\documentclass[12pt,a4paper]{article}
\usepackage{amsfonts}

\usepackage[english]{babel}
\usepackage[latin1]{inputenc}
\usepackage[OT1]{fontenc}
\usepackage{amsmath,amssymb, amsthm}
\usepackage[pdftex]{graphicx}
\usepackage{verbatim}
\usepackage{float}
\usepackage{placeins}
\usepackage{flafter}
\usepackage{longtable}
\usepackage{rotating}
\usepackage{hhline}
\usepackage{eurosym}
\usepackage{natbib}
\usepackage[onehalfspacing]{setspace}
\usepackage{lscape}
\usepackage[usenames]{color}
\usepackage{indentfirst}
\usepackage{url}
\usepackage[top=1in,bottom=1in,left=1in,right=1in,nohead]{geometry}
\usepackage{bigstrut,multirow,bigdelim}
\usepackage{subfloat}
\usepackage[position=above]{subfig}
\usepackage{etoolbox}
\usepackage{mathtools}
\usepackage{bm}
\usepackage{booktabs}
\usepackage{enumerate}

\input{tcilatex}

\newtheorem{assumption}[theorem]{Assumption}

\newcommand{\eps}{\varepsilon}
\DeclarePairedDelimiter\abs{\lvert}{\rvert}
\DeclarePairedDelimiter\norm{\lVert}{\rVert}
        
\newcommand{\addnote}[1]{\begin{spacing}{1}\footnotesize{\noindent #1}\end{spacing}}

\usepackage{array}
\newcolumntype{C}[1]{>{\centering\let\newline\\\arraybackslash\hspace{0pt}}m{#1}}

\begin{document}

\title{Estimation of a Heterogeneous Demand Function with~Berkson~Errors}
\author{Richard Blundell\thanks{
Department of Economics, University College London (UCL), and Institute for
Fiscal Studies (IFS). 7 Ridgmount Street, London WC1E 7AE, United Kingdom,
email: r.blundell@ucl.ac.uk. }, Joel Horowitz\thanks{%
Department of Economics, Northwestern University, and Cemmap. 2211 Campus Drive, Evanston, Illinois 60208, USA, email: joel-horowitz@northwestern.edu. }
\ and Matthias Parey\thanks{%
School of Economics, University of Surrey, and Institute for Fiscal
Studies (IFS). Guildford, Surrey GU2 7XH, United Kingdom, email:
m.parey@surrey.ac.uk.} }
\date{August 2019\thanks{%
We thank seminar participants at the `Heterogeneity in Supply and Demand'
conference at Boston College and the Bristol Econometric Study Group Annual
Conference for helpful comments. We are grateful to Agnes Norris Keiller for invaluable help in assembling the  gasoline price data. We would also like to thank the ESRC Centre for the
Microeconomic Analysis of Public Policy at IFS and the ERC advanced grant MICROCONLAB
at UCL for financial support.}}
\maketitle

\begin{abstract}
\thispagestyle{empty}%

\singlespacing
\noindent Berkson errors are commonplace in empirical microeconomics.  In consumer demand this form of measurement error occurs when the price an individual pays is measured by the (weighted) average price paid by individuals in a specified group (e.g., a county), rather than the true transaction price. We show the importance of such measurement errors for the estimation of demand in  a setting with nonseparable unobserved heterogeneity. We develop a consistent estimator using external information on the true  distribution of prices. Examining the demand for gasoline in the U.S., we document substantial within-market price variability, and show that there are significant spatial differences in the magnitude of Berkson errors across regions of the U.S. Accounting for Berkson errors is found to be quantitatively important for  estimating price effects and for welfare calculations. Imposing the Slutsky shape constraint greatly reduces the sensitivity  to Berkson errors.

\bigskip

\noindent \textbf{JEL:}\ C14, C21, D12

\bigskip

\noindent \textbf{Keywords:} consumer demand, nonseparable models, quantile
regression, measurement error, gasoline demand, Berkson errors.
\end{abstract}

\thispagestyle{empty}%

\newpage

\section{Introduction}

\setcounter{page}{1}%

Datasets that are commonly used in microeconometric work often suffer from a particular type of measurement error in
the covariates: Instead of observing the true covariate a household faces, the
researcher observes a group-level (weighted) average, such as a regional average (e.g.\ in a county). The resulting errors in the covariate are called \textit{Berkson errors}. Berkson measurement errors occur frequently in applied econometric analyses in which information on relevant covariates is not collected directly from households in a survey, but is taken from an alternative data source and assigned to households based on their location or other characteristics. While covariates
assigned in this way will often be highly correlated with the true
covariates, they will not be identical as long as there is some variability
in the covariate within the specified group. %
For example, in the gasoline demand application discussed in Sections 5-6 of this paper, counties experience within-county price variability of up to 10 percent around the mean, and within-county variation accounts for a substantial share of the overall variation in prices. Furthermore, the amount of within-county price variability differs substantially across regions of the U.S., as shown in Figure  \ref{fig:plot_map_sd} below. 
%
%
\begin{figure}[t!h]
\caption{Within-market variability in gas prices across counties}
\label{fig:plot_map_sd}%
\vspace{-7mm}
\begin{center}
{\fbox{%
\includegraphics [width=\textwidth]{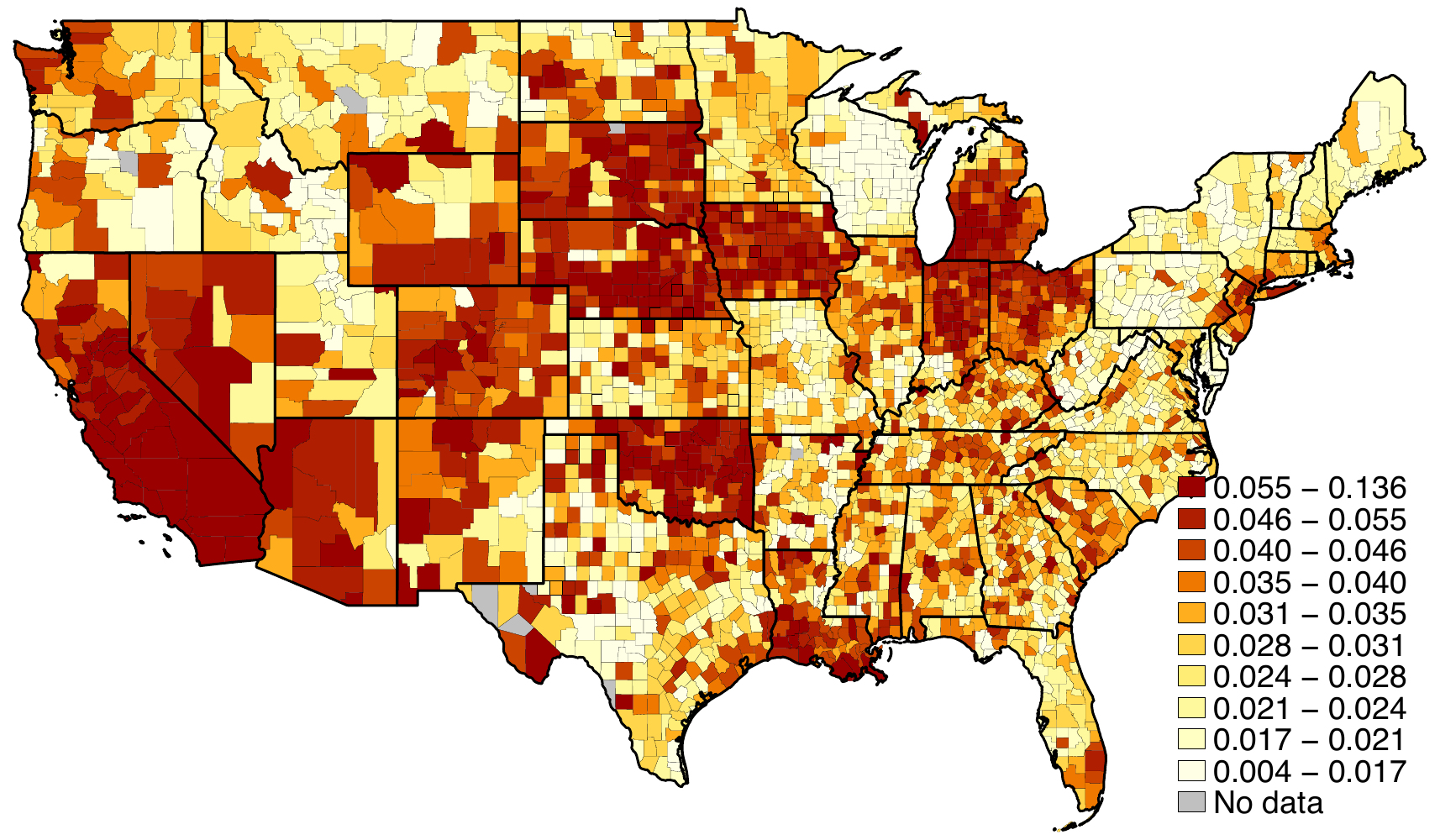}}}
\end{center}
\vspace{-2mm}%
{\footnotesize {Note: Map shows county-level standard-deviation in (log) gasoline prices (after removing county effects and day fixed effects). Based on station-level data from \citet{yilmazkuday-2017}, see Section \ref{sec:dispersion} below for details.} }
\end{figure}
%

Textbook analysis of this
kind of econometric model often focuses on the case when the model is linear in the
covariate and the error is additive. In this case, Berkson errors do not
lead to a bias. This is sometimes taken to mean that Berkson errors are
unlikely to cause significant bias in applied analysis, compared to say
classical measurement error. However, these results no longer hold when the model is nonlinear. In nonlinear models, Berkson errors are not innocuous and require careful treatment.

In this paper we consider estimation of a demand model with nonseparable unobserved heterogeneity with Berkson errors. Consider the demand function with nonseparable unobserved heterogeneity 
\begin{equation}\label{eqn:g_noberkson}
Q = G(P,Y,U)
\end{equation}
where $Q$ is the quantity demanded, $P$ the price, $Y$ household income, and 
$U$ unobserved heterogeneity. Suppose now that we do not observe the true price at which a transaction took place, which we
refer to as $P^\star$. Instead, we observe a county average price $P$ that
is related to $P^\star$ by 
\begin{equation}
P^\star =P+\epsilon,
\end{equation}
where $\epsilon$ is an unobserved random variable, independent of $P$. With these Berkson errors \citep%
{berkson-1950}, the demand model becomes 
\begin{equation}\label{eqn:g_berkson}
Q = G(P+\epsilon ,Y,U).
\end{equation}
Importantly, Berkson errors in variables are different from classical errors
in variables, where $P = P^\star + \epsilon$, with $\epsilon$ independent of 
$P^\star$.

In this paper, we argue that understanding the role of Berkson measurement errors in demand estimation is of growing relevance. The focus on understanding heterogeneity in responses motivates researchers to investigate behavior at different points in the distribution of unobserved heterogeneity, see e.g.\ \citet{Browning-Carro2007}. Moreover, researchers are increasingly interested in nonlinear models with non-separable unoberserved heterogeneity, see e.g. references in \citet{Cameron-Trivedi2005, bhp-2012, bhp-2017}. Better data and increased computational power facilitate the study of models that do not impose linearity restrictions and, instead, allow flexible functional forms with a high degree of potential nonlinearity. Accordingly, nonlinear models are increasingly important in applications. 

This paper develops a method for estimating a nonseparable demand model in
the presence of Berkson errors, using a Maximum Likelihood Estimator (MLE) of quantiles of demand conditional on price and income.
The standard quantile approach is inconsistent when prices are subject to Berkson errors. The maximum likelihood procedure we propose estimates all quantiles simultaneously,
and
a monotonicity constraint is used to ensure that the estimated quantiles do
not cross. This estimator enables us to contrast the resulting estimates to results
obtained assuming the absence of Berkson errors. Our estimation procedure accounts for spatial differences in the extent of Berkson error across locations, a feature which we find to be quantitatively important in our empirical application.

\cite{delaigle-hall-qiu-2006} show the demand function is unidentified nonparametrically unless either the
distribution of  the Berkson error is known or can be estimated consistently from
auxiliary data. Alternatively identification can be delivered if  there is
an
instrument that is related to the true price in a suitable
way
\citep{schennach-2013}. We choose to follow the first of these
approaches and use auxiliary data from external sources to inform us about the distribution of
the Berkson error.
 We then assess the sensitivity to Berkson errors across different levels of the Berkson error variance. Finally, we note there is a potential for prices to be endogenous. To address this we develop a test
for the exogeneity of covariates in the presence of Berkson errors.

We motivate and illustrate our analysis with an application to gasoline
demand. Household travel surveys frequently assign gasoline prices from
external sources based on the location of the household, leading to the
presence of Berkson errors. A long-standing body of work has documented the
importance of allowing for potential non-linearities in household gasoline
demand (\cite{hausman-newey-1995, yatchew-no-2001, bhp-2012}). The role of
unobserved heterogeneity motivates a quantile modelling approach (\cite%
{bhp-2017,hoderlein-vanhems-2018}). These considerations suggest that
nonlinearity plays an important role in this appliciation, highlighting the importance of Berkson errors in applied research and the need to treat them carefully.

We find that accounting for Berkson errors is quantitatively important. For
example, Deadweight Loss measures derived from our estimates differ
substantially when we allow for Berkson errors. In previous work we have
investigated the role of shape restrictions in semiparametric or
nonparametric estimation settings (\cite{bhp-2012,bhp-2017}). In a setting
with Berkson errors, we find that imposing shape restrictions, in the form of the Slutsky inequality, reduces the
sensitivity of the estimates to the presence of Berkson errors.

The paper proceeds as follows. In the next section, we introduce Berkson errors and outline the demand
model. Section 3 develops the MLE estimator.
Section 4 presents the exogeneity test. In Section 5, we describe household gasoline data, and also document external evidence on the distribution of current local gasoline prices.  
The estimation results for the gasoline demand responses to prices and for deadweight loss welfare measures are presented in Section 6. Section 7 concludes.

\section{Berkson Errors and the Demand Model}

We begin this section by providing examples of where Berkson errors occur in applied microeconometric work. We then focus on Berkson errors in demand analysis; we outline the nonseparable demand model in the absence of
Berkson errors, and then introduce Berkson errors into the model.

\subsection{Examples of Berkson Errors}

Berkson errors occur commonly in applied econometric work. Our application is to prices in consumer demand. Here we describe three additional  examples where Berkson errors are likely to be relevant.

A common case is a situation where the researcher does not observe the true value of the variable of interest, but instead only observes an indicator for the group the individual belongs to. The researcher then assigns a `typical value' from an external dataset, often the \textit{group average}. This group will often be a geographic identifier or a time period. For example, relevant covariates may not be surveyed or measured at the level of the household, but are instead approximated by a regional average from an external source.  
For example, \citet{currie-neidell-2005} study the effect of air pollution on infant death in California. Since pollution measures are not available at the household level, exposure is measured for each zip code as a weighted average of all fixed pollution monitors installed within a given distance from the zip code centroid. 

Another case is the situation where implementing a treatment exposes individuals to unobserved \textit{heterogeneity in the treatment intensity}. In this setting, the intensity of a treatment varies randomly and in unobserved manner across treated units. This could be due to variability in the technology delivering the intervention, or due to differences in the staff implementing the treatment. For example, the dose of a drug delivered might vary slightly across patients, the amount of fertilizer spread on plots might vary randomly, or the support provided to unemployed workers might vary with the caseworker. For the latter case, \citet{schiprowski-2019} documents significant variation in the effectiveness of caseworker meetings for unemployed workers, depending on the productivity of the caseworker; at the same time, details of the caseworker assigned to the unemployed are typically not observed by researchers.

A third case is a situation where individuals are uncertain about a relevant quantity and provide an \textit{optimal prediction} \citep{hyslop-imbens-2001}. For example, respondents in a survey are asked about a quantity they are uncertain about, and provide the best estimate of this quantity given their information set. When respondents provide an optimal prediction, the resulting prediction error is uncorrelated to the reported value. \citet{phelps-1972} develops a model where firms infer productivity from a noisy signal as well as characteristics of the individual, and form an optimal prediction by a weighted average of the signal and the expected value given the characteristics. Due to the optimal prediction any deviation from the true value will be unrelated to the prediction made by the uncertain individual.

Survey data frequently asks respondents to provide details on variables where respondents may be uncertain about the exact values. For example, \citet{chan-stevens-2004} investigate how pension accumulations affect retirement decisions. Since data on pensions is self-reported in their data, the authors consider the possibility that the pension measure may be a noisy measure of the truth, predicted by the survey respondent, leading to Berkson error.    

In another setting, \citet{hastings-kane-staiger-2009} study benefits from attending higher achieving schools. They use Bayes' rule to infer a parental preference parameter from a model of demand for schools, and in turn estimate models which allow the benefits of attendance to vary with this parental preference parameter. The measurement error in the estimated parental preference parameter then exhibits Berkson errors.

In all these cases, the variable used in the analysis differs from the variable which is relevant for the outcome in a way described by the Berkson error framework. This underscores the wide relevance of these kinds of measurement error for applied econometric work.

\subsection{The Demand Model and the Presence of Berkson Errors}

\label{the-model}

In the absence of Berkson errors, the demand function with nonseparable unobserved heterogeneity is set out in equation (\ref{eqn:g_noberkson}) in Section 1.
We assume for now that $U$ is a scalar random variable that is statistically
independent of $(P,Y)$, and that $G(P,Y,U)$ is monotone increasing in its
third argument.\footnote{%
The assumption of scalar unobserved heterogeneity ($U$) is restrictive but
necessary to achieve point identification and to do welfare analysis. Section 4 treats the possibility that P is endogenous and, therefore, that $U$ and $P$ are correlated. \cite%
{hausman-newey-2017} and \cite{dette-hoderlein-neumeyer-2016}  discuss
models with multi-dimensional unobserved heterogeneity.} We further assume
without further loss of generality that $U \sim \text{U}[0,1]$. 

Under these
assumptions, the $\alpha$ quantile of $Q$ conditional on $(P,Y)$ is 
\begin{equation*}
Q_\alpha = G(P,Y,\alpha) \equiv G_\alpha (P,Y).
\end{equation*}
That is, the the conditional $\alpha$ quantile of $Q$ recovers the demand
function $G$, evaluated at $U=\alpha$.

With Berkson errors, the demand model becomes equation (\ref{eqn:g_berkson}). The function $G$ is unidentified nonparametrically unless either the
distribution of $\epsilon$ is known or can be estimated consistently from
auxiliary data (\cite{delaigle-hall-qiu-2006}) or, alternatively, there is an
instrument $Z$ that is related to the true price $P^\star$ in a suitable way
(\cite{schennach-2013}). In this work we follow the first of these
approaches, and use auxiliary data to inform us about the distribution of
the Berkson error.

\section{Estimation}

\label{section:MLE}

\subsection{A Maximum Likelihood Estimator}

In this section we develop the Maximum Likelihood Estimation approach. The
model is 
\begin{equation*}
Q=G(P+\epsilon,Y,U); \;\; U \sim \text{U}[0,1].
\end{equation*}
Therefore,  
\begin{align}
P(Q \leq z | P,Y ) & = P(G(P+\epsilon,Y,U)\leq z | P,Y ) = P( U \leq
G^{-1}(P+\epsilon,Y,z) | P,Y )  \label{eqn1} \\
& = \int G^{-1}(P + \epsilon ,Y , z) f_\epsilon (\epsilon )d\epsilon  \notag
\\
& = E_\epsilon G^{-1}(P+\epsilon ,Y,z),  \notag
\end{align}
where $G^{-1}(\cdot,\cdot,z)$ is the inverse of $G$ in the third argument.

The left-hand term of equation (\ref{eqn1}), $P(Q \leq z | P,Y)$, is
identified by the sampling process. $G^{-1}$ and $G$ are identified
nonparametrically if and only if $G^{-1}$ is determined uniquely by 
\begin{equation*}
P(Q \leq z | P,Y ) = E_\epsilon G^{-1}(P + \epsilon ,Y , z ).
\end{equation*}
This requires knowledge of $f_\epsilon(\epsilon)$; \cite%
{delaigle-hall-qiu-2006} present a similar identification result for a
conditional mean model.\footnote{Note that the identification condition can be
formulated as a version of the completeness condition of Nonparametric
Instrumental Variables (NPIV) models. See \citet{Newey-powell-2003}.}

The truncated series  
\begin{align}
G^{-1}(P+\epsilon ,Y,Q) \approx \sum_{j=1}^{J} \theta_j\Psi_j (P+\epsilon
,Y,Q)
\end{align}
provides a flexible parametric approximation to $G^{-1}$. In the truncated series, $J$ is the (fixed) truncation point, the $\Psi_j$ 's are basis functions and the $\theta_j$ 's are Fourier
coefficients. The data  ${\{Q_i, P_i, Y_i:  i=1,...,n\}}$ are a random sample of \(n\) households. The log-likelihood function for estimating parameter vector $%
\theta$ is the logarithm of the probability density of the data.  This is:
\begin{equation*}
\log L(\theta)=\sum_{i=1}^n log\sum_{j=1}^{J_n} \theta_j \int \left. \frac{%
\partial \Psi_j(P_i+\epsilon ,Y_i,z)}{\partial z} \right|_{z=Q_i} f_\epsilon
(\epsilon )d\epsilon
\end{equation*}
Maximum likelihood estimation of $\theta$ consists of maximizing $\log
L(\theta)$ subject to the following constraints: first, that $G^{-1}$ is
non-decreasing in its third argument, and second, $0 \leq G^{-1} \leq 1$.
The maximum likelihood procedure estimates all quantiles simultaneously, and
by imposing the monotonicity constraint above ensures that the estimated quantiles do
not cross. For the presentation of the results, we numerically invert the
estimated function $\hat{G}^{-1}$ to obtain the corresponding demand
function $\hat{G}$.

\subsection{Shape Restrictions}
In some of the estimates we also impose the Slutsky shape restriction from consumer theory.
Assuming quantity, income and prices for household $i$ are measured in logs,
and $S_i$ reflects the budget share of household $i$, the Slutsky
constraint, evaluated at $(P_i, Y_i, U_i)$ can be written as 
\begin{align*}
\frac{\partial Q}{\partial P}(P_i,Y_i,U_i) + \frac{\partial Q}{\partial Y}%
(P_i,Y_i,U_i) \; S_i \leq 0.
\end{align*}
From $U=G^{-1}(P,Y,Q)$, we re-write the price and income effect in terms of $%
G^{-1}$, so that the Slutsky condition for household $i$ is 
\begin{align}
\frac{\partial G^{-1}}{\partial P}(P_i,Y_i,Q_i) + \frac{\partial G^{-1}}{%
\partial Y}(P_i,Y_i,Q_i) \; S_i \geq 0.  \label{eqn-slutsky}
\end{align}
The estimation then proceeds by maximizing the log-likehood as before,
adding the constraint (\ref{eqn-slutsky}) for a set of households in the
data. 

\section{An Exogeneity Test}

A common concern in demand estimation is the possible endogeneity of the price
variable, where local prices are correlated with consumer preferences (see 
\cite{bhp-2012,bhp-2017}). If a variable $W$ is available as an instrument
for the price, the researcher can test for the presence of endogeneity. In a nonparametric or flexible parametric model, such a test is likely to have better power properties than a comparison of the
exogenous estimate with an instrumental variables (IV) estimate. We
therefore develop an exogeneity test, which takes account of the presence of
Berkson errors. In this section we state the test statistic and asymptotic approximation to its distribution. The corresponding derivations can be found in
Appendix \ref{section:appendix_exog_test}.

Assume that the instrument, $W$, satisfies
\begin{align*}
        P(U\leq \tau | W,Y)=\tau.
\end{align*}
Let $G^{-1}_{EX}$ denote the inverse demand function $G^{-1}$, described
in Section 3, under the null hypothesis $H_0$ that $P$ is exogenous. Under
$H_0$
\begin{align}
        & \text{Pr}\left[G_{EX}^{-1}(P+\eps,Y,Q| W=w, Y=y) \leq \tau \right]
\nonumber \\
        & = E \int I\left[ G_{EX}^{-1} (P+\eps, Y,Q | W=w, Y=y) \leq \tau
\right] f_\eps(\eps) d\eps = \tau \label{eqn-nr-2}
\end{align}
for any $(y,w)$ in the support of $(Y,W)$. The exogeneity test statistic
is based on a sample analog of this relation. Let $f_{YW}$ denote the probability
density function of $(Y,W)$. Let $K$ be a probability density function that
is supported on $[-1,1]$ and symmetrical around 0. Let
$\{h_n: n=1,2,...\}$ be a sequence of positive numbers that converges to
0 as $n \rightarrow \infty$. $K$ is called a kernel function and $\{h_n\}$
is called a sequence of bandwidths. Denote the data by $\{Q_i,P_i,Y_i,W_i:
i=1,\dots,n\}$. Let $\hat{f}_{YW}$ be a kernel nonparametric estimator of
$f_{YW}$:
\begin{align*}
        \hat{f}_{YW}(y,w) = \frac{1}{nh_n^2} \sum_{i=1}^n K\left(\frac{W_i-w}{h_n}\right)K\left(\frac{Y_i-y}{h_n}
\right).
\end{align*}
Let $\hat{G}^{-1}_{EX}$ denote the MLE of ${G}^{-1}_{EX}$. Define
\begin{align*}
        S_n(y,w) = & \frac{1}{nh^2}\sum_{i=1}^n\left\{ \int I\left[\hat G_{EX}^{-1}(P_i
+ \varepsilon, Y_i, Q_i) \leq \tau\right] f_\varepsilon (\varepsilon) d\varepsilon
        K\left(\frac{W_i-w}{h_n}\right)K\left(\frac{Y_i-y}{h_n}\right)\right\}.
\end{align*}
$S_n(y,w)/\hat{f}_{YW}(y,w)$ is a sample analog of the integral expression
in (\ref{eqn-nr-2}). The test statistic is
\begin{align*}
        T_n = nh_n^2 \int \left[ S_n(y,w) - \tau \hat{f}_{YW}(y,w) \right]^2
dwdy.
\end{align*}
To obtain an asymptotic approximation to the distribution of $T_n$, assume
without loss of generality that $(y,w) \in [0,1]^2$. Let $\{\hat\lambda_j:
j=1,\dots,n\}$ denote the eigenvalues of the operator
\begin{align*}
        C(y_1,w_1; y_2,w_2) = \tau(1-\tau) \hat{f}_{YW}(y_1,w_1) \int K(\xi)
K(\xi + \delta_W) K(\zeta) K(\zeta + \delta_Y) d\xi d\zeta.
\end{align*}
Let $\{L_n: n=1,2,\dots\}$ be an increasing sequence of positive constants
such that $L_n \rightarrow \infty$ and $n^{-1/2} L_n^{3/2} \rightarrow 0$
as $n \rightarrow \infty$. Under regularity conditions that are stated in
the appendix,
\begin{align*}
\left| T_n - \sum_{j=1}^{L_j} \hat{\lambda}_j \chi^2_j \right| \rightarrow^p
0
\end{align*}
as $n\rightarrow \infty$, where the $\chi_j^2\;$s are independent random
variables that are distributed as chi-square with one degree of freedom.
The distribution of $T_n$ can be approximated by that of
\begin{align*}
\omega=\sum_{j=1}^{L_n} \hat\lambda_j \chi_j^2.
\end{align*}
The quantiles of the distribution of $\omega$ can be estimated with any desired
accuracy by Monte Carlo simulation.

\section{Data on Demand and Prices}

\subsection{The household gasoline demand}

The data are from the 2001 National Household Travel Survey (NHTS), which
surveys the civilian noninstitutionalized population in the United States.
This is a household level survey conducted by telephone, and complemented by
travel diaries and odometer readings.\footnote{%
See \cite{ornl-2004} and \cite{bhp-2012} for further detail on the survey.}
These data provide information on the travel behavior of selected
households. We focus on annual mileage by vehicles owned by the
household.

In order to minimize heterogeneity in the sample, the following restrictions
are imposed: We restrict attention to households with a white respondent,
two or more adults, and at least one child under age 16. We drop households
in the most rural areas, where farming activities are likely to be
particularly important. We also omit households in Hawaii due to its
different geographic situation compared to the continental states.
Households without any drivers or where key variables are not observed are
excluded, and we restrict attention to gasoline-based vehicles (excluding
diesel, natural gas, or electricity based vehicles).\footnote{%
We require gasoline demand of at least one gallon, and we drop one outlier
observation where the reported gasoline share is larger than 1.} The sample
we use is the same as in \cite{bhp-2017}.

A key aspect of the data is that although  odometer readings
and  fuel efficiencies are recorded,  price information is not collected at the household level,
reflecting the expense in collecting purchase diaries and the resulting
burden for respondents (\cite{eia-2003,eia-2011}). Instead, in the NHTS
gasoline prices are assigned the fuel cost in the local area, based on the
location of the household \citep{eia-2003}. In Section \ref{sec:dispersion}
we document that households face substantial price variability within local markets, and
we use this information to assess the extent of Berkson errors.

The resulting sample contains 3,640 observations. Table 1 presents
summary statistics. 
%
\begin{table}[ht]
\caption{Sample descriptives}
\vspace{-7mm}
\label{table:sample}
\begin{center}
\begin{tabular}{lcc}
\hline
\hline
      & \textit{Mean} & \textit{St.\ dev.} \bigstrut\\
\hline
      &       &  \bigstrut[t]\\
Log gasoline demand & 7.127 & 0.646 \\
Log price & 0.286 & 0.057 \\
Log income & 11.054 & 0.580 \\
      &       &  \bigstrut[b]\\
\hline
Observations & \multicolumn{2}{c}{3640} \bigstrut\\
\hline
\end{tabular}%

\end{center}
\par
{\addnote{Note: Table presents mean and standard deviations. See text for details.} }
\end{table}
%
The reported means of our key variables correspond to
about 1,250 gallons of gasoline per year, a gasoline price of \$1.33, and
household income of about \$63,000. For reference, Table 2 presents baseline
estimates of price and income elasticities from a log-log model of gasoline
demand. 
%
\begin{table}[hbt]
\caption{Log-log model estimates}
\label{table:loglog}
\vspace{-7mm}
\begin{center}
\begin{tabular}{cccccc}
\hline
\hline
      & $\alpha=0.25$ & $\alpha=0.50$ & $\alpha=0.75$ &       & OLS \bigstrut[t]\\
      & (1)   & (2)   & (3)   &       & (4) \bigstrut[b]\\
\hline
      &       &       &       &       &  \bigstrut[t]\\
$\log(p)$ & -1.00 & -0.72 & -0.60 &       & -0.83 \\
      & [0.22] & [0.19] & [0.22] &       & [0.18] \\
      &       &       &       &       &  \\
$\log(y)$ & 0.41  & 0.33  & 0.23  &       & 0.34 \\
      & [0.02] & [0.02] & [0.02] &       & [0.02] \\
      &       &       &       &       &  \\
Constant & 2.58  & 3.74  & 5.15  &       & 3.62 \\
      & [0.25] & [0.21] & [0.25] &       & [0.20] \\
\hline
$N$     & 3640  & 3640  & 3640  &       & 3640 \bigstrut\\
\hline
\end{tabular}%

\end{center}
\par
{\addnote{Note: Dependent variable is log gasoline demand. See text
for details.} }
\end{table}
%
In the mean regression model, we find a price elasticity of -0.83
and an income elasticity of 0.34, similar to the elasticities reported in
other studies of gasoline demand (see further \cite{bhp-2017}). Looking
across quantiles, we find the lower quantile households to be more sensitive
to changes in prices and income.

In the estimation below, the function $G^{-1}$ is specified as a product of three Chebyshev polynomials, one each for $P$, $Y$, and $Q$.  We use cubic polynomials in
price and income, and a 7th-degree polynomial in quantity. The high-degree polynomial in quantity enables us to estimate differences in the demand function across quantiles of the distribution of unobserved heterogeneity.\footnote{We also trim the top and bottom 1 percent of the quantity distribution.} 
When we impose the Slutsky constraint, using the
observed data points in the sample, we restrict attention to those data
points broadly in the areas of the data which we are focusing our analysis
on below.\footnote{%
For this purpose, we add restrictions for data points between the 10th and
the 90th percentile of the unconditional demand data, 0.2 to 0.36 in the log
price dimension, and household income between 20,000 and 90,000 USD.}

\subsection{Dispersion in local gasoline prices}

\label{sec:dispersion}

In this subsection, we present evidence on the within-market dispersion of
gasoline prices. To gain insight into this, we draw on data collected by \citet{yilmazkuday-2017}, containing daily gas prices for virtually all gas station in the U.S. during a one-month period (July 2015) from MapQuest (\url{http://gasprices.
mapquest.com}).\footnote{We have also collected data on local gas price variability from \url{www.gasbuddy.com} for a set of seven examplary counties in the US. The within-county variability from these GasBuddy prices is very similar to the estimate from the MapQuest data that we describe in this section; we focus on the MapQuest data due to its almost universal coverage.} Since these data are based on fleet transaction data, they are likely to be highly accurate. Together with the almost universal coverage of gas stations in these data, this dataset is very well suited for our purpose (see \citet{yilmazkuday-2017} for further detail on these data).\footnote{We exclude Alaska and Hawaii from the subsequent analysis to focus on the contiguous United States. Gas stations are assigned to counties based on their zip code.}  

\begin{figure}[t!h]
\caption{Within-market price distribution}
\label{fig:plot_histogram_within}%
\vspace{-7mm}
\begin{center}
{\fbox{%
\includegraphics [width=110mm]{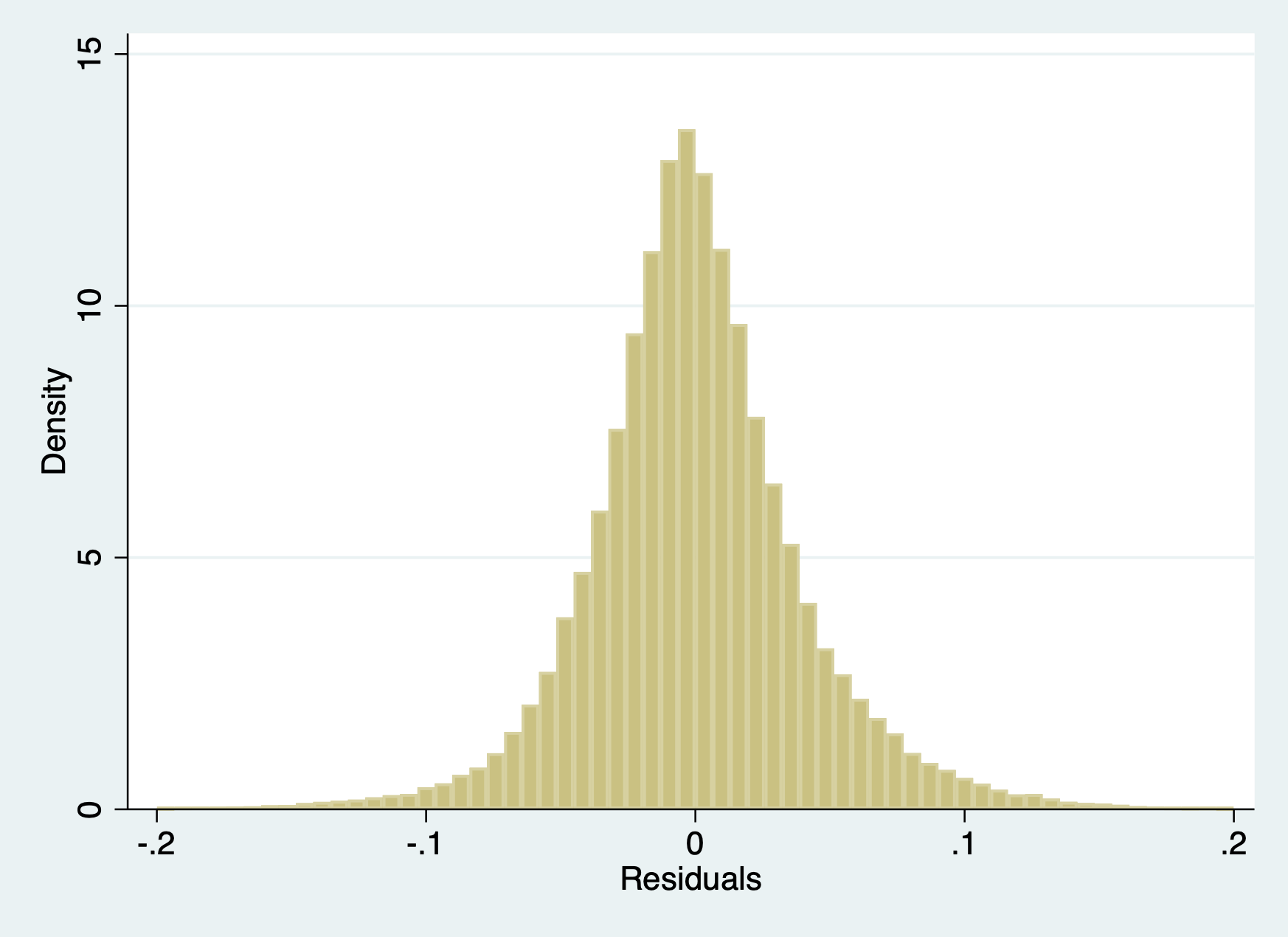}}}
\end{center}
\vspace{-2mm}%
{\footnotesize {Note: Histogram shows distribution of gasoline prices, after removing county effects and day fixed effects. See text for details.} }
\end{figure}

To provide a description of the within-market price variability, Figure \ref{fig:plot_histogram_within}
shows a histogram of the gas prices (measured in logs), after removing county fixed
effects and day fixed effects. 
%
%
This histogram shows that there is substantial within-market price variability. Within the same county (and having accounted for day effects), prices vary frequently by up to 10 percent in either direction. The histogram also suggests that a normal approximation of the within-market
dispersion broadly captures the shape of the distribution. 

\begin{figure}[t!h]
\caption{Variability in price distribution across counties}
\label{fig:mapquest_across}%
\vspace{-7mm}
\begin{center}
{\fbox{%
\includegraphics [width=110mm]{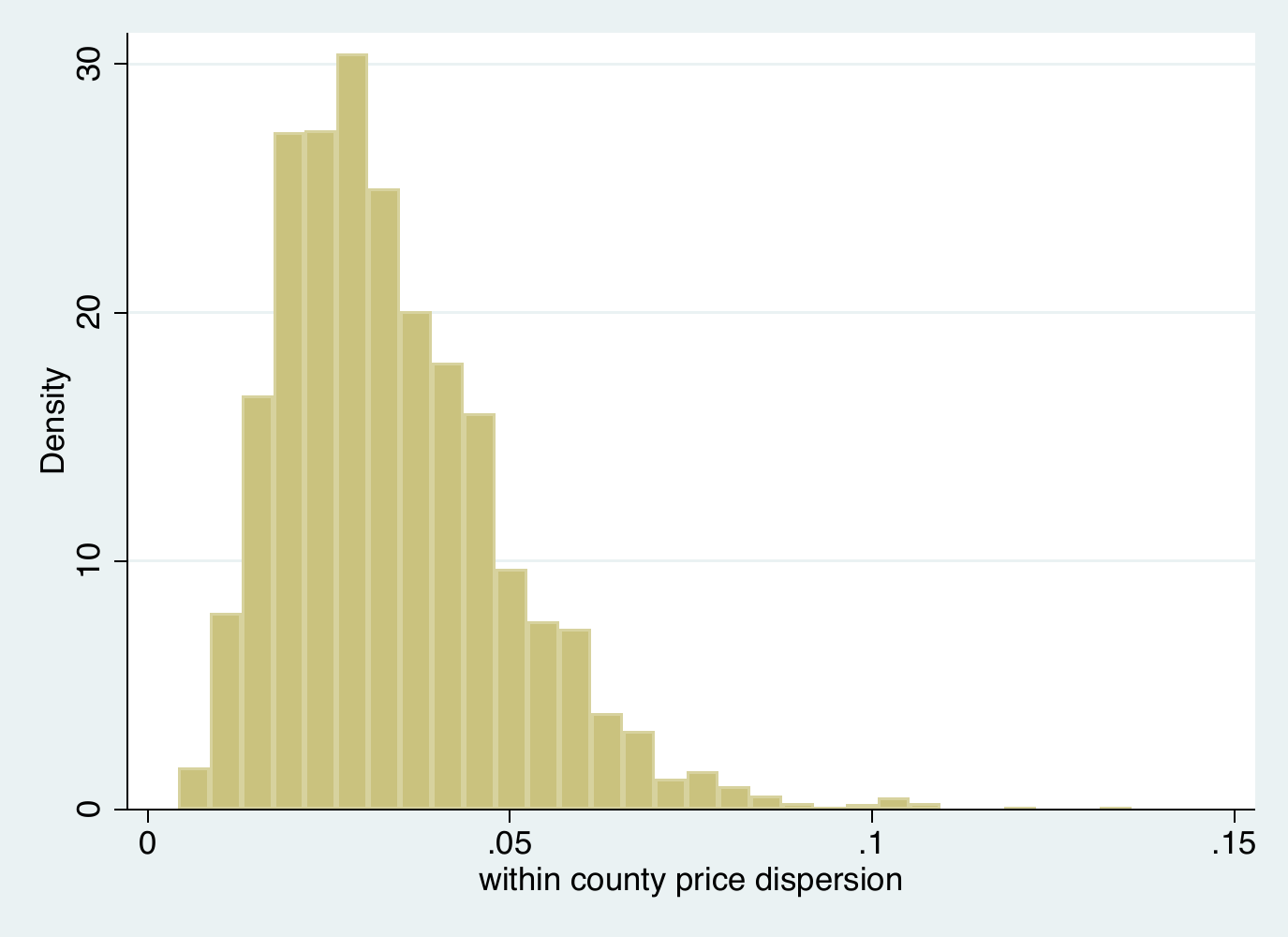}}}
\end{center}
\vspace{-2mm}%
{\footnotesize {Note: Histogram shows distribution of county-level gas price dispersion, measured as standard deviation of (log) gasoline prices (after removing county effects and day fixed effects). See text for details.} }
\end{figure}

To show the variability across counties in the price dispersion, we compute the within-county price dispersion for each county in the US. The resulting map is shown in Figure \ref{fig:plot_map_sd} above.
As is evident from the map, price dispersion varies across the United States. For example, price variability is particularly high in California, but also in other states, such as Oklahoma, South Dakota or Nevada. Figure \ref{fig:mapquest_across} shows the histogram of this within-county price dispersion. 
%
%
Across all counties, the mean value (unweighted) of the within-market dispersion is 0.0339 (with first and third quartile taking values of 0.022 and 0.042).  
Comparing this value to the reported standard deviation of 0.057 in the NHTS
price variable (see Table 1) shows that a significant amount of price
variability occurs within local markets, suggesting that the Berkson error
is an important feature of the price variation in this sample. In our empirical analysis, we use a normal distribution for the Berkson error, and allow the standard deviation to vary by U.S.\ state. To specify the standard deviation, we use a weighted average across the counties in each state (see Figure \ref{fig:plot_map_sd}). This accounts for the substantial differences in the amount of Berkson errors across different parts of the U.S., and incorporates the spatial pattern of Berkson errors into our estimation. 

\subsection{Gasoline price cost shifter}

To examine the exogeneity of prices we  require a variable which is correlated with
gasoline prices, but uncorrelated with the unobservable type of the
household. Building on earlier work \citep{bhp-2012}, we use
transportation cost as a cost shifter. This reflects that the cost of
transporting the fuel from the supply source is an important determinant of
prices. 

We measure transportation cost with the distance between one of the
major oil platforms in the Gulf of Mexico and the state capital. The U.S. Gulf Coast region accounts for the majority of total
U.S. refinery net production of finished motor gasoline and for almost
two-thirds of U.S. crude oil imports. It is also the starting point for most
major gasoline pipelines. We therefore expect that transportation cost
increases with distance to the Gulf of Mexico (see Blundell et al., 2012,
for further details and references). Appendix Figure \ref{fig:plot_iv} shows the
systematic and positive relationship between state-level average prices and
the distance to the Gulf of Mexico.

\section{Empirical Results}

\subsection{Demand estimates}
%
\begin{figure}[tpbh!]
                \caption{MLE estimates at the median (at middle income)}
\vspace{-7mm}
                \begin{center}
\subfloat{\includegraphics[trim=28mm 92mm 3cm 88mm, clip=true,width=0.75\textwidth]{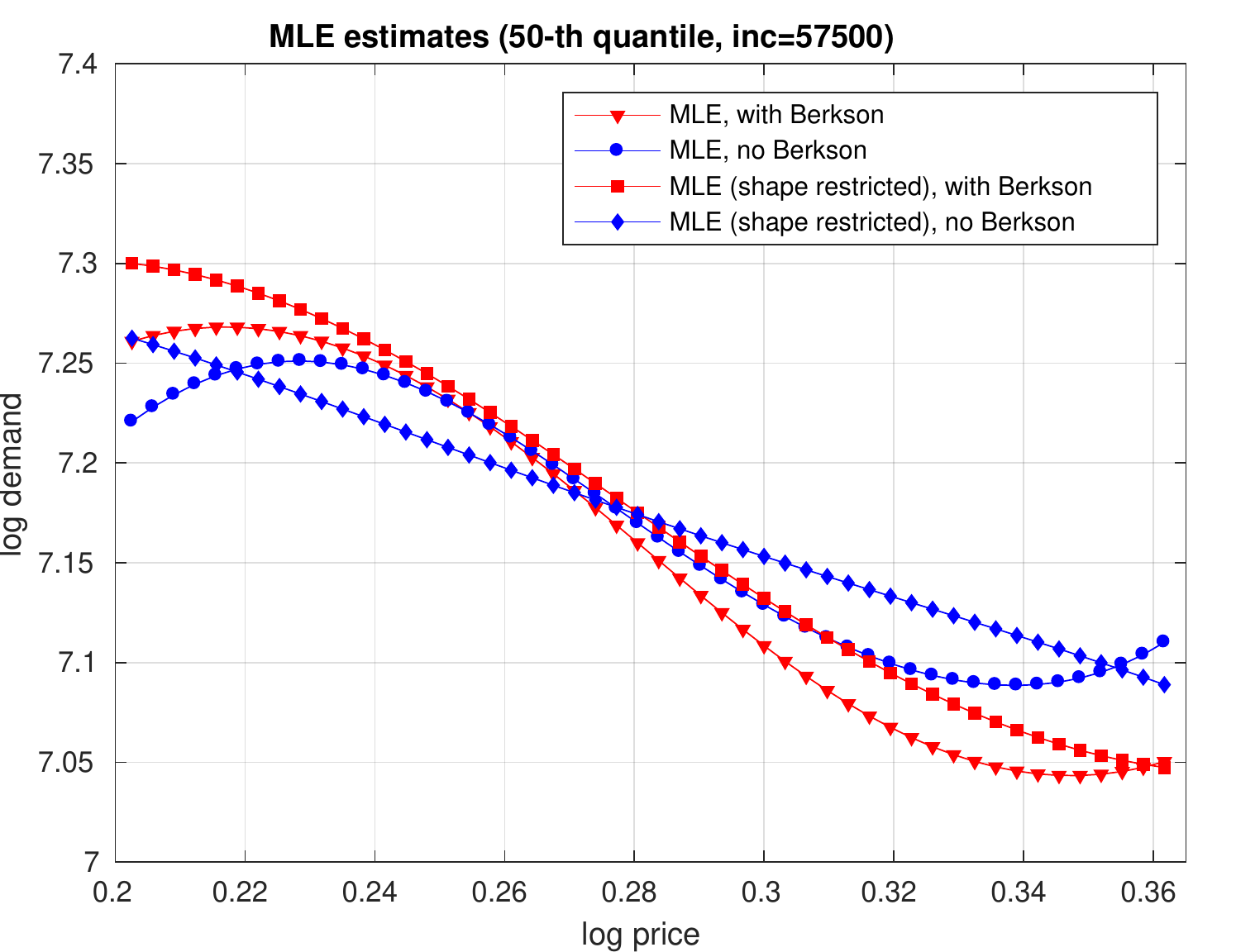}}
\end{center}
\par
{\addnote{Note: The figure shows MLE estimates at the median ($\tau=0.50$) for the middle income group. Lines shown in red are estimates accounting for Berkson error, lines shown in blue assume absence of Berkson error. The figure compares unconstrained estimates versus Slutsky-constrained estimates (see legend). See text for details.} }

\label{fig:main}
\end{figure}
\begin{figure}[pbh!]
                \caption{MLE estimates across quartiles (at middle income)}
\vspace{-7mm}
                \begin{center}
\subfloat{\includegraphics[trim=30mm 35mm 3cm 18mm, clip=true,width=0.75\textwidth]{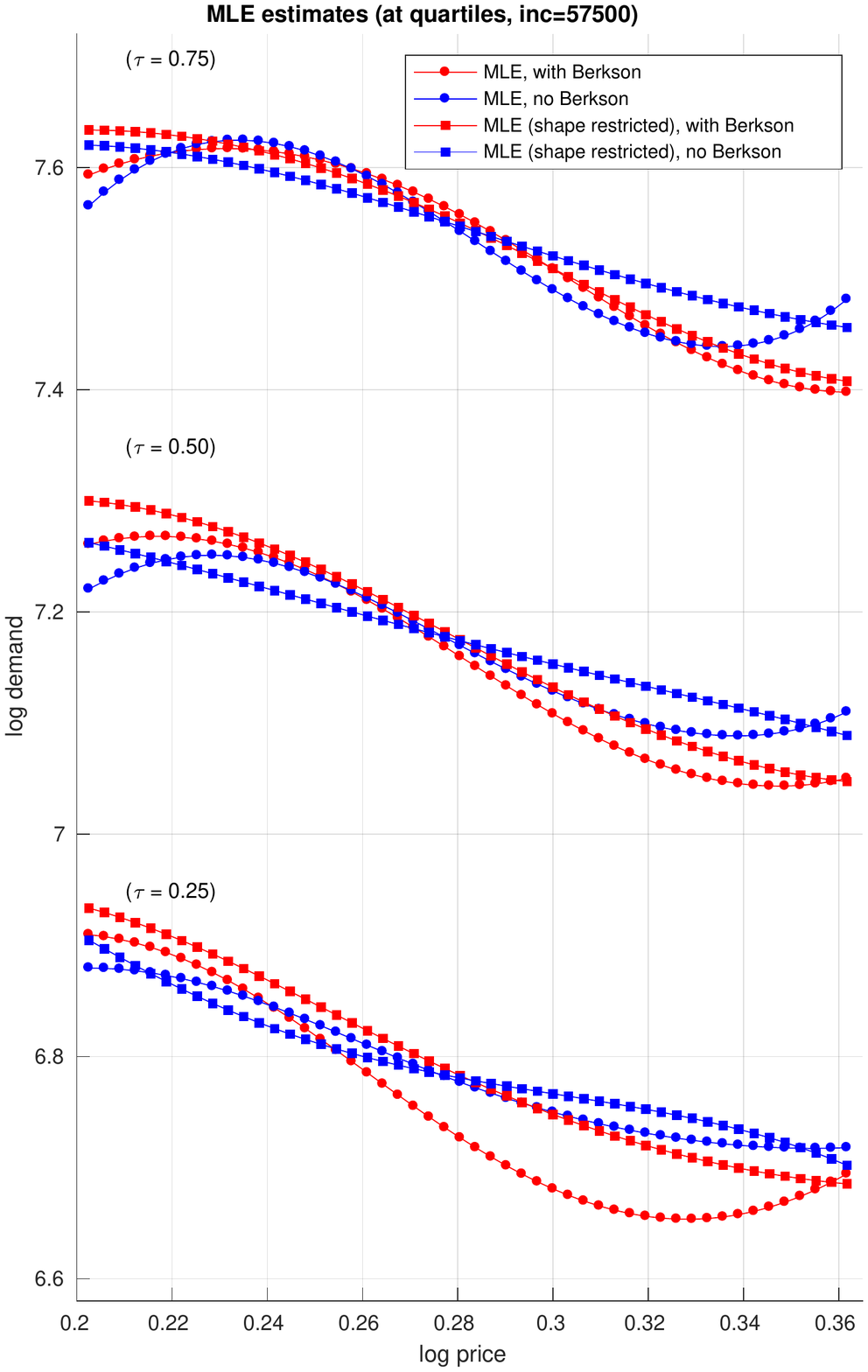}}
\end{center}
\par
{\addnote{Note: The figure shows MLE estimates at the three quartiles (upper quartile, $\tau=0.75$, median, $\tau=0.50$, and lower quartile, $\tau=0.25$) for the middle income group. Lines shown in red are estimates accounting for Berkson error, lines shown in blue assume absence of Berkson error.  The figure compares unconstrained estimates versus Slutsky-constrained estimates (see legend). See text for details.} }

\label{fig:comparison_quantiles}
\end{figure}

%
Figure \ref{fig:main} shows the ML estimates for the median, for the middle income group (\$57,500). Figure \ref{fig:comparison_quantiles} compares the estimates across the quartiles of the distribution of the
unobserved heterogeneity, for the same income group.
%
%
The 
round markers show the MLE estimates without taking account of Berkson errors; the upside down triangular markers
show the MLE with Berkson error. As can be seen from the Figure, accounting
for Berkson errors accentuates the variability in the demand estimates, and
leads to relevant differences in the estimated price responsiveness. For the
median, for example, shifting the price across the full range shown in the
figure (from 0.20 to 0.36) leads to a fall in estimated (log) demand by 0.11
assuming the absence of Berkson errors, compared to 0.21 in the presence of
Berkson errors. Note the non-monotonicity in the unconstrained demand curve estimates, which is an artifact of random sampling variation (see further \citet{bhp-2012,bhp-2017}). This non-monotonicity appears to accentuate the sensitivity to the Berkson errors in this empirical example.

   The square markers in Figures \ref{fig:main} and \ref{fig:comparison_quantiles} show the estimates when we impose Slutsky
negativity. Although there is still a difference in the slope, the two sets
of estimates are now much more similar.  Looking across the different quantiles, we note  a consistent finding that imposing the Slutsky inequality restriction removes non-monotonicity and delivers a smoother estimated demand curve much less sensitive to Berkson errors. This suggests that the estimates
under the shape restriction are less sensitive to accounting for Berkson
errors, reflecting the stabilizing effect of the shape restriction on the
demand estimate. 

\begin{figure}[pbh!]
                \caption{MLE estimates across the income distribution (at $\tau=0.50$)}
                \begin{center}
                                        \vspace{-9mm}%
                \subfloat{\label{fig3a}\includegraphics[trim=30mm 92mm 3cm 88mm, clip=true,width=0.6\textwidth]{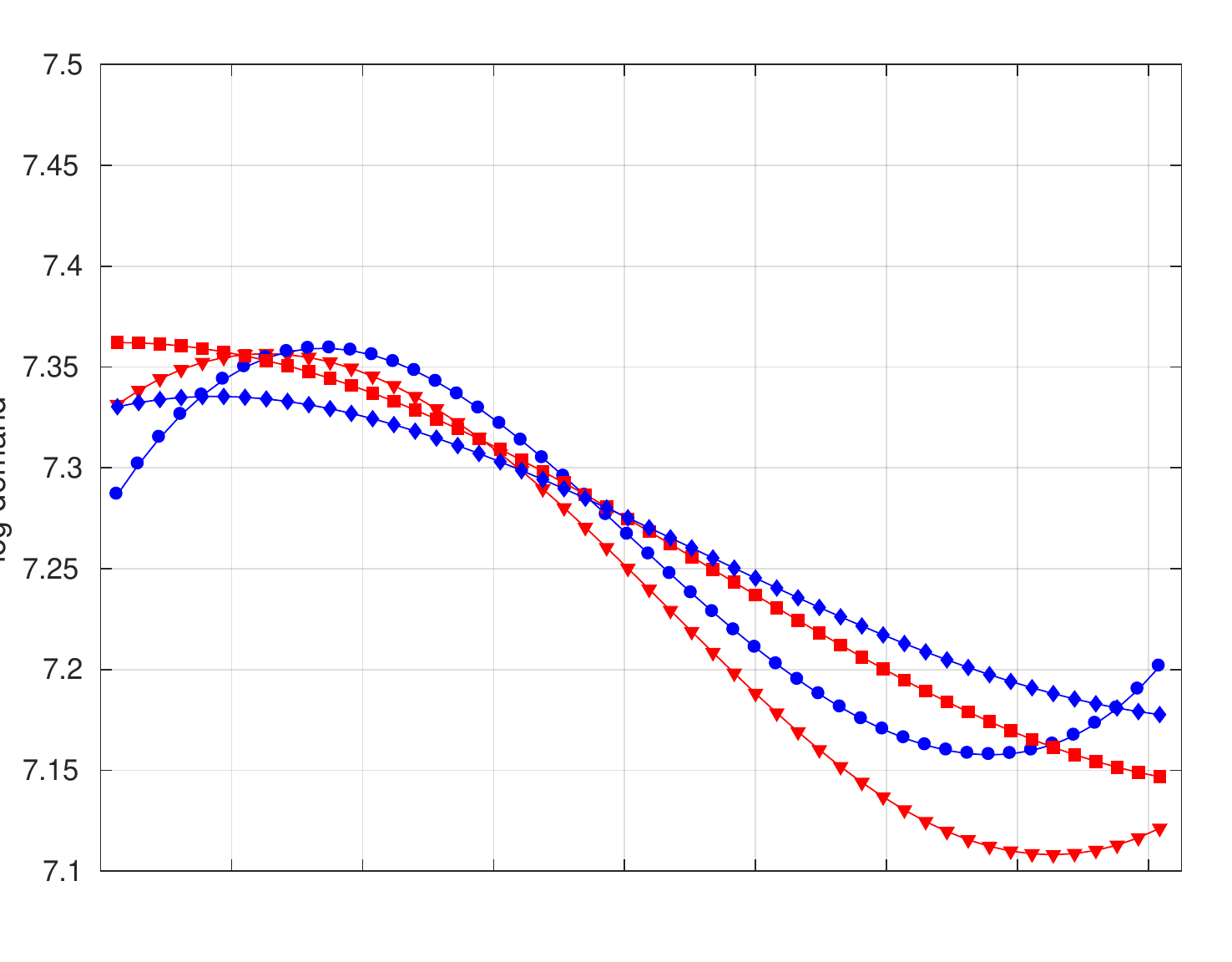}}\vspace{-12mm}\\
        \subfloat{\label{fig3b}\includegraphics[trim=30mm 92mm 3cm 88mm, clip=true,width=0.6\textwidth]{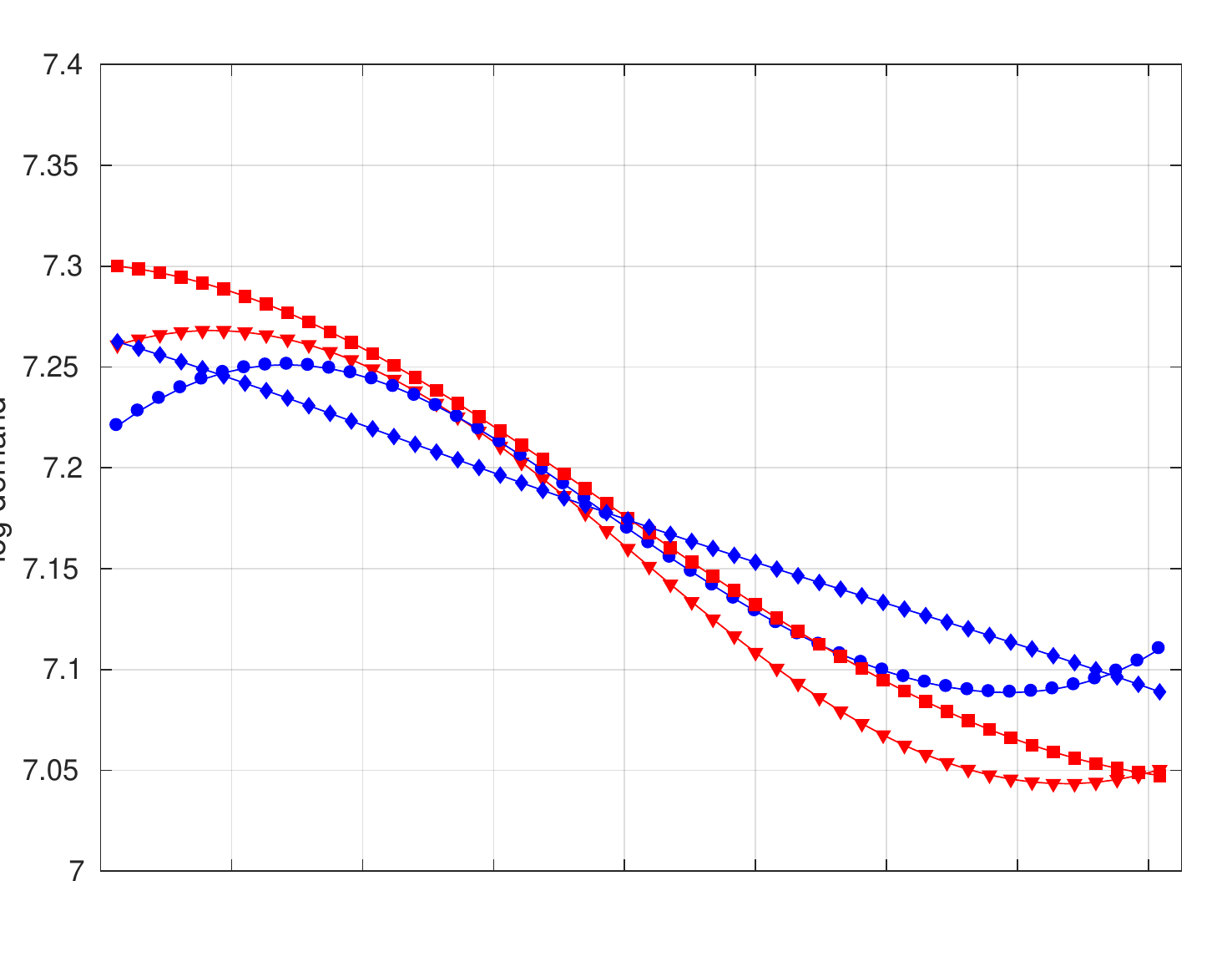}}\vspace{-12mm}\\
        \subfloat{\label{fig3c}\includegraphics[trim=30mm 90mm 3cm 88mm, clip=true,width=0.6\textwidth]{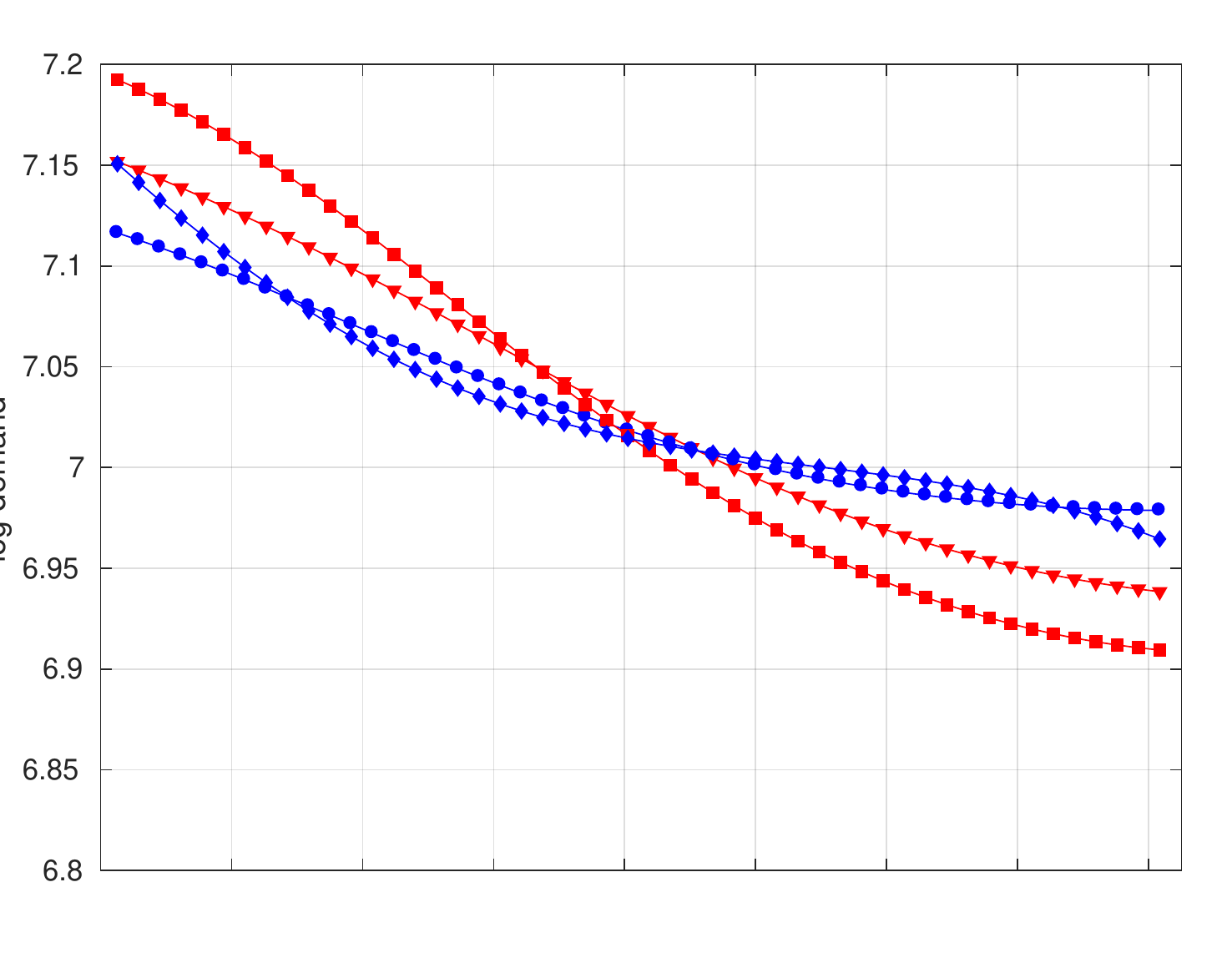}}\vspace{-10mm}\\
                \subfloat{\label{fig3d}\includegraphics[trim=71mm 130mm 71mm 130mm, clip=true,width=0.4\textwidth]{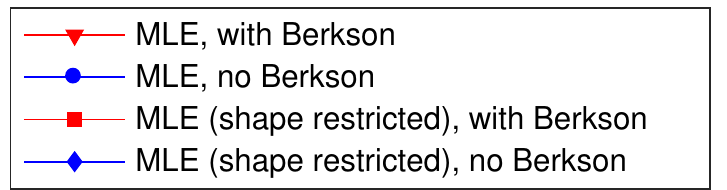}}\vspace{-8mm}\\ 
        \end{center}
\par
{\addnote{Note: The figure shows MLE estimates for the three income groups (top panel: `high' income, corresponding to \$72,500, middle panel: `medium' income, corresponding to \$57,500, and bottom panel: `low' income, corresponding to \$42,500) at the median ($\tau=0.50$). Lines shown in red are estimates accounting for Berkson error, lines shown in blue assume absence of Berkson error. The figure compares unconstrained estimates versus Slutsky-constrained estimates (see legend). See text for details.} }
        \label{fig:comparison_y}
        \end{figure}

Figure \ref{fig:comparison_y} compares the estimated effect at the median across
the income distribution, comparing \mbox{\$72,500}, \mbox{\$57,500}, and \mbox{\$42,500},
representing upper, middle and lower income households, respectively.  These results  highlight the importance of the Slutsky restriction in achieving monotonicity. In this way, these results not only provide demand function estimates that are consistent with consumer theory, but in addition attenuate sensitivity to Berkson errors. However, although the mitigation of sensitivity to Berkson errors through imposing the Slutsky restriction is a clear empirical finding of our analysis, we do not claim that it is a theoretical necessity.   

%

\begin{figure}[pbh!]
                \caption{Comparison of different magnitudes of the Berkson error}
                \begin{center}
                        \subfloat[factor 1]{\label{fig4a}\includegraphics[trim=30mm 92mm 3cm 88mm, clip=true,width=0.5\textwidth]{fig_g_both_tau50_nhts2001_state_ff1_ally_p3_q7_y3_yy57500.pdf}}\\
        \subfloat[factor 0.8]{\label{fig4b}\includegraphics[trim=30mm 92mm 3cm 88mm, clip=true,width=0.5\textwidth]{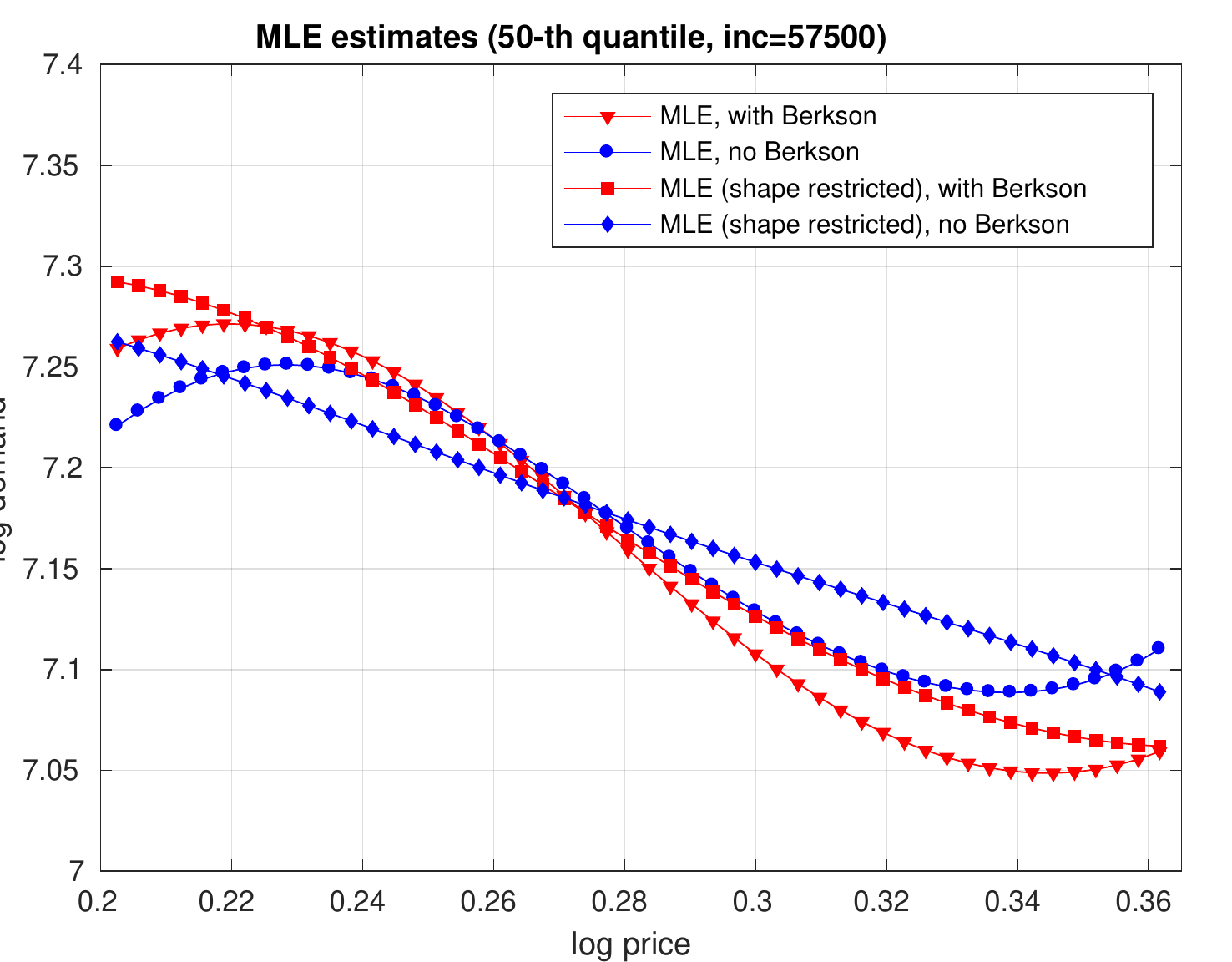}}\\
        \subfloat[factor 1.2]{\label{fig4c}\includegraphics[trim=30mm 92mm 3cm 88mm, clip=true,width=0.5\textwidth]{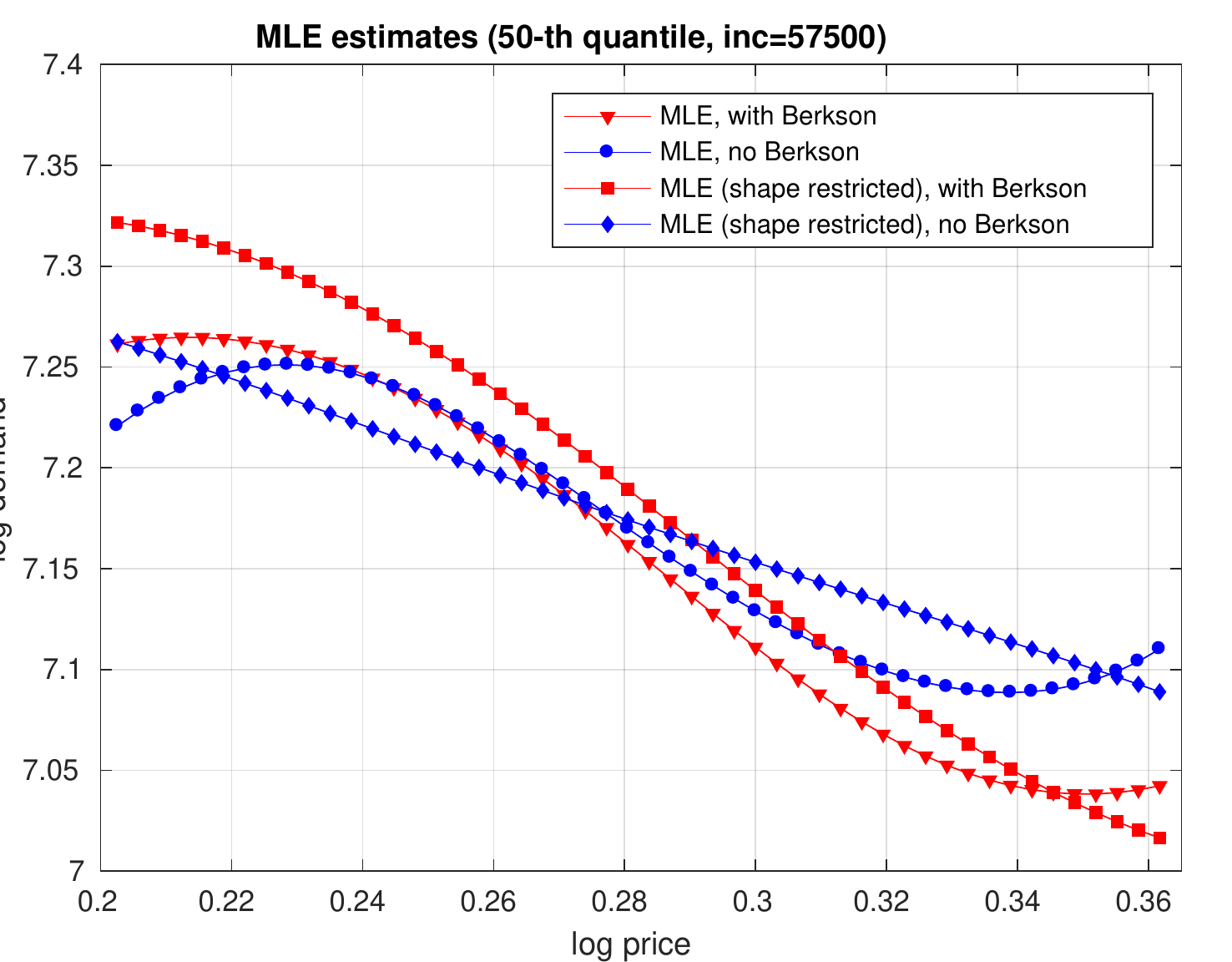}}\\
        \end{center}
\par
{\addnote{Note: The figure compares the baseline estimates in panel (a) to estimates with different standard deviation of the Berkson error. Panel (b) reduces the Berkson error standard deviation by factor 0.8, and panel (c) increases it by factor 1.2. Estimates shown for the median, at the middle income group. Lines shown in red are estimates accounting for Berkson error, lines shown in blue assume absence of Berkson error. Round markers indicate unconstrained estimates, square markers indicate Slutsky-constrained estimates. See text for details.} }

        \label{fig:sensitivity_berkson}
        \end{figure}

Figure \ref{fig:sensitivity_berkson} compares the estimates for different magnitudes of the Berkson
error, varying the standard deviation with factor 1.2 and factor 0.8,
respectively.
For small standard deviations (panel (b)), the presence of
Berkson error makes very little difference to the demand estimates. However
for larger standard deviation of the Berkson errors (panel (c)), the
differences become quantitatively very important. This is especially
pronounced for the unconstrained estimates.
%
%

\subsection{Estimating the welfare loss of gasoline taxation}

The estimates of the demand function can be used to estimate welfare
measures such as deadweight loss (DWL). We consider a hypothetical tax
change which moves the price from $p^{0}$ to $p^{1}$ in a discrete fashion
(see \cite{bhp-2017}). Let $e(p)$ denote the expenditure function at price $p
$ and a reference utility level. The DWL of this price change is then given
by 
\begin{equation*}
L(p^{0},p^{1})=e(p^{1})-e(p^{0})-(p^{1}-p^{0})\;H_{\alpha }\left[
p^{1},e(p^{1})\right] ,
\end{equation*}%
where ${H}_{\alpha}(p,y)$ is the Marshallian demand function. $L(p^{0},p^{1})
$ is computed by replacing $e$ and $H$ with consistent estimates. The
estimator of $e$, $\hat{e}$, is constructed by numerical solution of the
differential equation 
\begin{equation*}
\frac{d\hat{e}(t)}{dt}=\hat{H}_{\alpha }\left[ p(t),\hat{e}(t)\right] \frac{%
dp(t)}{dt},
\end{equation*}%
where $\left[ p(t),\hat{e}(t)\right] $ ($0\leq t\leq 1$) is a
price-(estimated) expenditure path.

Deadweight Loss (DWL) estimates are reported in Table \ref{table:dwl}.
%
\begin{table}[hbt]
\caption{DWL estimates}
\label{table:dwl}
\begin{center}
\begin{tabular}{cccccccc}
\toprule
\toprule
      &       &       & \multicolumn{2}{c}{with Berkson errors} &       & \multicolumn{2}{c}{without Berkson errors} \\
\cmidrule{4-5}\cmidrule{7-8}      &       &       &       &       &       &       &  \\
      &       &       & \multicolumn{1}{C{4.085em}}{DWL per } & \multicolumn{1}{C{4.085em}}{DWL per } &       & \multicolumn{1}{C{4.085em}}{DWL per } & \multicolumn{1}{C{4.085em}}{DWL per } \\
      & income &       & \multicolumn{1}{C{4.085em}}{tax} & \multicolumn{1}{C{4.085em}}{income} &       & \multicolumn{1}{C{4.085em}}{tax} & \multicolumn{1}{C{4.085em}}{income} \\
\midrule
      &       &       & \multicolumn{1}{C{4.085em}}{(1)} & \multicolumn{1}{C{4.085em}}{(2)} &       & \multicolumn{1}{C{4.085em}}{(3)} & \multicolumn{1}{C{4.085em}}{(4)} \\
\midrule
      &       &       & \multicolumn{2}{c}{} &       & \multicolumn{2}{c}{} \\
\multicolumn{8}{l}{\boldmath{}\textbf{A. Upper quartile ($\tau$=0.75)}\unboldmath{}} \\
      &       &       &       &       &       &       &  \\
      & high  &       & 0.155 & 7.82  &       & 0.054 & 3.00 \\
\textit{unconstrained} & middle &       & 0.146 & 8.80  &       & 0.055 & 3.59 \\
      & low   &       & 0.116 & 8.70  &       & 0.043 & 3.34 \\
      &       &       &       &       &       &       &  \\
      & high  &       & 0.116 & 6.18  &       & 0.094 & 5.14 \\
\textit{constrained} & middle &       & 0.140 & 8.52  &       & 0.093 & 5.92 \\
      & low   &       & 0.165 & 11.65 &       & 0.065 & 5.03 \\
      &       &       &       &       &       &       &  \\
\multicolumn{8}{l}{\boldmath{}\textbf{B. Median ($\tau$=0.50)}\unboldmath{}} \\
      &       &       &       &       &       &       &  \\
      & high  &       & 0.130 & 4.70  &       & 0.061 & 2.40 \\
\textit{unconstrained} & middle &       & 0.117 & 4.96  &       & 0.062 & 2.80 \\
      & low   &       & 0.101 & 5.17  &       & 0.052 & 2.80 \\
      &       &       &       &       &       &       &  \\
      & high  &       & 0.130 & 4.82  &       & 0.096 & 3.66 \\
\textit{constrained} & middle &       & 0.139 & 5.90  &       & 0.092 & 4.04 \\
      & low   &       & 0.133 & 6.62  &       & 0.069 & 3.66 \\
      &       &       &       &       &       &       &  \\
\multicolumn{8}{l}{\boldmath{}\textbf{C. Lower quartile ($\tau$=0.25)}\unboldmath{}} \\
      &       &       &       &       &       &       &  \\
      & high  &       & 0.087 & 2.20  &       & 0.077 & 2.03 \\
\textit{unconstrained} & middle &       & 0.067 & 1.98  &       & 0.074 & 2.24 \\
      & low   &       & 0.064 & 2.28  &       & 0.069 & 2.50 \\
      &       &       &         &       &       &       &  \\
      & high  &       & 0.139 & 3.51  &       & 0.102 & 2.64 \\
\textit{constrained} & middle &       & 0.118 & 3.48  &       & 0.094 & 2.80 \\
      & low   &       & 0.087 & 3.07  &       & 0.083 & 2.96 \\
\bottomrule
\bottomrule
\end{tabular}%

\end{center}
\par
{\addnote{Note: DWL shown corresponds to a price change from the 5th to the 95th percentile in the data. Income level `high' corresponds to \$72,500, `medium' to \$57,500, and `low' to \$42,500. `DWL per income' is re-scaled by $\times 10^4$ for readibility.} }
\end{table}
%
Looking at the
unconstrained estimates, the table shows the strong quantitative difference
in the DWL figures between the estimates with Berkson error (columns
(1)-(2)) versus those without (columns (3)-(4)). In many cases, the
estimates with Berkson errors but not the Slutsky restriction are more than twice as large as those assuming
absence of Berkson errors.

Regarding the constrained estimates, however, the DWL figures are now much
closer together and often of similar order of magnitude. This underlines a key
point from the demand curve  estimates in the previous subsection, the Slutsky constrained demand estimates
reduce sensitivity to the presence of Berkson errors.

\subsection{Exogeneity test}

In this section we report the empirical results for the endogeneity test. To
simplify the computation, we implement the univariate version of the test and specify a common standard deviation of the Berkson error distribution across the U.S.\footnote{We set the standard deviation to 0.033, which is the (unweighted) mean across counties in the U.S., see further Section \ref{sec:dispersion} above.}
For this purpose, we stratify the sample along the income dimension in three
groups: a low-income group of households (household income between \$35,000
and \$50,000), a middle-income group of households (between \$50,000 and
\$65,000), and an upper-income group of households (between \$65,000 and
\$80,000). The test is then performed for each income group. The results are shown in Table \ref{table:exog}. 

We find we do not reject exogeneity for any of the three income groups. This conclusion remains unchanged when we consider moderate variation in the extent of the Berkson error, multiplying the standard error of the Berkson error by a factor of 0.8 and 1.2, respectively, as shown in the table. The critical values shown in the table do not take account of the fact that we perform the test three times (for each of the three income groups). One possibility for adjusting the size for a joint 0.05 level test would be a Bonferroni adjustment. The adjusted $p$-value for a joint 0.05 level
test of exogeneity is $1 - (0.95)^{(1/3)} = 0.01695$, at each of the three
income groups. Using this more conservative cutoff would strengthen our conclusion. Based on these results, endogeneity
is unlikely to be a first-order issue for our estimates.
%
\begin{table}[hbt]
\caption{Exogeneity test}
\label{table:exog}
\begin{center}
\begin{tabular}{ccccc}
\toprule
\toprule
      & test statistic & crit value (5\%) & p-value & reject? \\
\midrule
      &       &       &       &  \\
\multicolumn{1}{l}{(a) HIGH INCOME (N=578)} &       &          &       &   \\
      &       &       &       &  \\
baseline case & 0.1575 & 0.4000 & 0.4490 & no \\
reduced Berkson error, factor 0.8 & 0.1629 & 0.4000 & 0.4291 & no \\
increased Berkson error, factor 1.2 & 0.1443 & 0.4000 & 0.5009 & no \\
      &       &       &       &  \\
\midrule
      &       &       &       &  \\
\multicolumn{1}{l}{(b) MEDIUM INCOME (N=555)} &       &       &       &  \\
      &       &       &       &  \\
baseline case & 0.2257 & 0.4033 & 0.2459 & no \\
reduced Berkson error, factor 0.8 & 0.1879 & 0.4033 & 0.3444 & no \\
increased Berkson error, factor 1.2 & 0.2617 & 0.4033 & 0.1781 & no \\
      &       &       &       &  \\
\midrule
      &       &       &       &   \\
\multicolumn{1}{l}{(c) LOW INCOME (N=580)} &       &       &       &   \\
      &       &       &       &  \\
baseline case & 0.1338 & 0.4042 & 0.5427 & no \\
reduced Berkson error, factor 0.8 & 0.1490 & 0.4042 & 0.4799 & no \\
increased Berkson error, factor 1.2 & 0.1777 & 0.4042 & 0.3768 & no \\
      &       &       &       &  \\
\bottomrule
\bottomrule
\end{tabular}%

\end{center}
\par
{\addnote{Note: Income range `high' refers to \$65,000-\$80,000, `medium' to \$50,000-\$65,000, `low' to \$35,000-\$50,000. Exogeneity test is conducted separately for each income range. Bonferroni-adjusted $p$-value for a joint 0.05 level test of exogeneity is $0.01695$. See text for details.} }
\end{table}
%

\section{Conclusions}

\label{conclusions}

It has long been understood that in a mean regression model with a linear
effect of a covariate with Berkson errors and an additive error term, the
coefficients in an OLS regression are unbiased. Recent advances in methods,
data, as well as computational capacity, together with a desire for understanding the effect of heterogeneity in the studied population,
have led to a growing interest in nonlinear models. In nonlinear models, the
role of Berkson errors is much less well understood, and ignoring these
errors in general leads to a bias in the estimates. This motivates our
interest in investigating the effect of Berkson errors, and methods for
addressing their presence in the data. We conduct this analysis in the
context of a quantile regression model, where the covariates enter through a
flexible parametric specification, allowing for potential nonlinearity in
the effects. Our application of interest is a gasoline demand model with
unobserved heterogeneity, where the price is measured with Berkson error.

The presence of Berkson errors is a frequent feature of economic data. It
occurs, for example, when the covariate is measured as a regionally
aggregated average, masking within-region variability. The data generating
process features the covariate which includes the Berkson error but its error-free value is
unobserved by the researcher. This naturally raises the question how much
difference recognizing the presence of Berkson error may make.

We derive a maximum likelihood estimator, which enables us to carry out
consistent estimation in the presence of Berkson errors with a known
density. The paper also develops a test for exogeneity of the Berkson
covariate in the presence of an instrument. 

We apply the method to the demand for gasoline in the U.S. We examine demand curves in which we impose the Slutsky inequality constraint and those that do not. The unconstrained  estimated
demand function display non-monotonicity in the price of gasoline. This estimated demand function is substantially affected by Berkson errors. The estimates
which do not take account of the Berkson errors understate the variability
in the price effect.  These results show that accounting for Berkson error
can have a substantial effect on the estimated demand function in a standard
demand application. In turn, these estimates result in differences in DWL
estimates for given price changes. In a number of cases, the DWL estimates recognizing the presence of Berkson errors are more than twice
as large as estimates assuming the absence of Berkson errors. Thus, Berkson errors can have quantitatively large effects.

In our application, the estimated demand function is 
weakly non-monotonic in the price. As Blundell et al. (2012, 2017) explain, this can be due to the effects of random sampling errors on the estimate. We overcome this problem by imposing the Slutsky constraint on the structural demand function estimates, as a way of adding structure
to the estimation problem. When the Slutsky restriction is imposed, the
estimated demand function is well-behaved and the effects of Berkson errors
are somewhat attenuated. These results illustrate that in a setting where
measurement error increases the uncertainty of the estimates, shape
restrictions such as the Slutsky constraint can be particularly useful for
providing additional structure to improve the estimation.

\bigskip \bigskip \bigskip

\clearpage

\setlength{\bibsep}{0pt plus 0.3ex} 
\bibliographystyle{biblio}
\bibliography{lit_Berkson}

\clearpage
\restoregeometry
\thispagestyle{empty}\appendix

\newgeometry{left=2cm,right=2cm,top=15mm,bottom=20mm} \thispagestyle{empty}%
\pagestyle{empty}

\section{Appendix}

\baselineskip 20pt \setcounter{figure}{0} \setcounter{table}{0} %
\renewcommand\thefigure{\thesection.\arabic{figure}} \renewcommand\thetable{%
\thesection.\arabic{table}}

\subsection{Additional Tables and Figures}

\begin{figure}[t!h]
\caption{Instrumental Variable for Price: Distance to the Gulf of Mexico}
\label{fig:plot_iv}%
\vspace{-7mm}
\begin{center}
{\fbox{%
\includegraphics [width=130mm, viewport= 0mm 0mm 170mm
120mm]{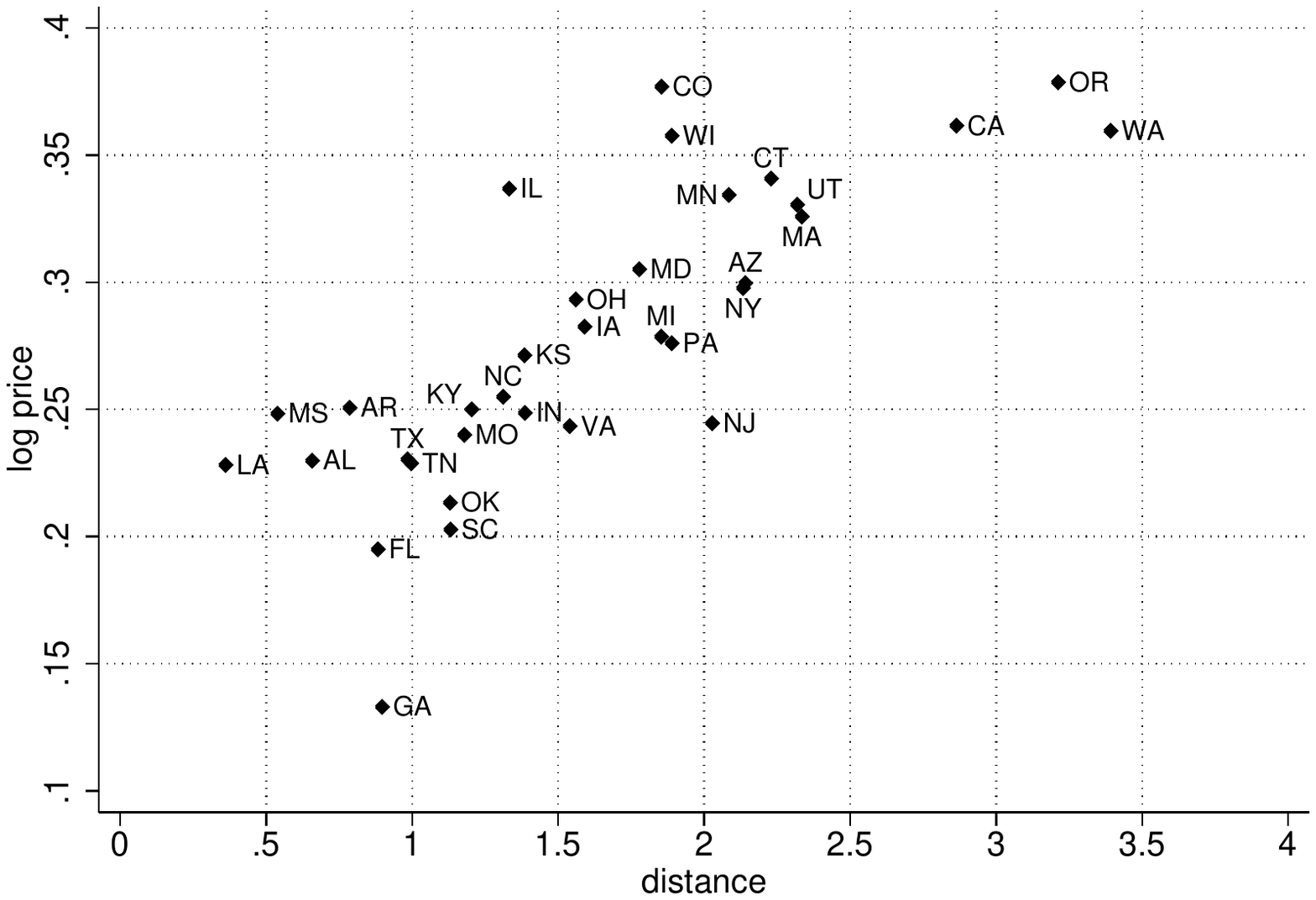}}}
\end{center}
\vspace{-2mm}%
{\footnotesize {Note: Price of gasoline and distance to the Gulf of Mexico. Distance to the respective state capital is measured in 1000 km. Source: BHP (2012, Figure 5).} }
\end{figure}

\begin{table}[hbt]
\caption{DWL estimates with confidence intervals}
\label{table:dwl-ci}
\begin{center}
\begin{tabular}{cccccccc}
\toprule
\toprule
      &       &       & \multicolumn{2}{c}{\textbf{with Berkson errors}} &       & \multicolumn{2}{c}{\textbf{without Berkson errors}} \\
\cmidrule{4-5}\cmidrule{7-8}      &       &       &       &       &       &       &  \\
      &       &       & \multicolumn{1}{C{6.75em}}{DWL per } & \multicolumn{1}{C{6.75em}}{DWL per } &       & \multicolumn{1}{C{6.75em}}{DWL per } & \multicolumn{1}{C{6.75em}}{DWL per } \\
      & income &       & \multicolumn{1}{C{6.75em}}{tax} & \multicolumn{1}{C{6.75em}}{income} &       & \multicolumn{1}{C{6.75em}}{tax} & \multicolumn{1}{C{6.75em}}{income} \\
\midrule
      &       &       & \multicolumn{1}{C{6.75em}}{(1)} & \multicolumn{1}{C{6.75em}}{(2)} &       & \multicolumn{1}{C{6.75em}}{(3)} & \multicolumn{1}{C{6.75em}}{(4)} \\
\midrule
      &       &       & \multicolumn{2}{c}{} &       & \multicolumn{2}{c}{} \\
\multicolumn{8}{l}{\boldmath{}\textbf{A. Upper quartile ($\tau$=0.75)}\unboldmath{}} \\
      &       &       &       &       &       &       &  \\
      & high  &       & 0.155 & 7.81  &       & 0.054 & 3.00 \\
      &       &       & [    0.101;     0.272] & [     5.44;     13.58] &       & [   -0.016;     0.103] & [    -0.51;      5.76] \\
      & middle &       & 0.146 & 8.80  &       & 0.055 & 3.59 \\
      &       &       & [    0.085;     0.238] & [     5.66;     13.99] &       & [   -0.005;     0.105] & [    -0.00;      6.87] \\
      & low   &       & 0.116 & 8.70  &       & 0.043 & 3.34 \\
      &       &       & [    0.009;     0.205] & [     1.74;     15.25] &       & [   -0.024;     0.123] & [    -1.51;      9.79] \\
      &       &       &       &       &       &       &  \\
\multicolumn{8}{l}{\boldmath{}\textbf{B. Median ($\tau$=0.50)}\unboldmath{}} \\
      &       &       &       &       &       &       &  \\
      & high  &       & 0.130 & 4.70  &       & 0.061 & 2.40 \\
      &       &       & [    0.046;     0.224] & [     1.99;      7.93] &       & [   -0.001;     0.114] & [     0.18;      4.44] \\
      & middle &       & 0.117 & 4.96  &       & 0.062 & 2.79 \\
      &       &       & [    0.045;     0.187] & [     2.22;      7.81] &       & [    0.010;     0.121] & [     0.56;      5.46] \\
      & low   &       & 0.101 & 5.17  &       & 0.052 & 2.80 \\
      &       &       & [   -0.005;     0.177] & [     0.40;      9.00] &       & [   -0.015;     0.132] & [    -0.52;      7.20] \\
      &       &       &       &       &       &       &  \\
\multicolumn{8}{l}{\boldmath{}\textbf{C. Lower quartile ($\tau$=0.25)}\unboldmath{}} \\
      &       &       &       &       &       &       &  \\
      & high  &       & 0.087 & 2.20  &       & 0.077 & 2.03 \\
      &       &       & [   -0.027;     0.170] & [    -0.41;      4.27] &       & [    0.012;     0.151] & [     0.48;      3.98] \\
      & middle &       & 0.067 & 1.98  &       & 0.074 & 2.24 \\
      &       &       & [   -0.031;     0.124] & [    -0.71;      3.66] &       & [    0.017;     0.147] & [     0.69;      4.48] \\
      & low   &       & 0.064 & 2.27  &       & 0.069 & 2.50 \\
      &       &       & [   -0.073;     0.138] & [    -2.03;      4.96] &       & [   -0.023;     0.157] & [    -0.36;      5.75] \\
      &       &       &       &       &       &       &  \\
\bottomrule
\bottomrule
\end{tabular}%

\end{center}
\par
{\addnote{Note: Table shows unconstrained DWL estimates with 90\% confidence intervals, based on 499 bootstrap replications. DWL shown corresponds to a price change from the 5th to the 95th percentile in the data. Income level `high' corresponds to \$72,500, `medium' to \$57,500, and `low' to \$42,500. `DWL per income' is re-scaled by $\times 10^4$ for readibility. See text for details.} }
\end{table}

\subsection{Exogeneity Test}

\label{section:appendix_exog_test}

The argument that follows uses linear functional notation. In this notation,
\begin{align*}
Pg=\int gdP; \;\;\; P_{n}g=\int gdP_n
\end{align*}
for any function $g(\cdot)$, where $P$ and $P_n$, respectively, are the
distribution and empirical distribution functions of the random argument of $g$.

To obtain an asymptotic approximation to the distribution of $T_n$, make:

\begin{assumption}
        \begin{enumerate}[(i)]
                \item $G^{-1}_{EX}$ is a known bounded function $g(\cdot,\cdot,\cdot,\theta)$, where $\theta \in \mathbb{R}^d$ for some $d<\infty$ is a constant parameter whose maximum likelihood estimate is denoted by $\hat\theta$ and whose true but unknown population value is denoted by $\theta_0$. 
                \item $n^{1/2}\left(\hat\theta - \theta_0\right) \rightarrow^d N(0,V)$ for some non-singular covariance matrix $V$.
                \item The first and second derivatives of $g$ with respect to its third argument are bounded and continuous uniformly over $\theta$ in a neighborhood of $\theta_0$ and the other arguments of $g$.
        \end{enumerate}
\end{assumption}

\begin{assumption}
        \begin{enumerate}[(i)]
                \item $K$ is a probability density function that is symmetrical about 0 and supported on $[-1,1]$.
                \item $n^{1/2}h/(\log n)^\gamma \rightarrow \infty$ as $n \rightarrow \infty$ for some $\gamma>1/2$.
        \end{enumerate}
\end{assumption}
Define
\begin{align*}
        G^{-1}_{EX}(\cdot,\cdot,\cdot)=g(\cdot,\cdot,\cdot,\theta).
\end{align*}
Define  
\begin{align*}
R_n(y,w,\varepsilon) = & \frac{1}{nh^2} \sum_{i=1}^n I\left[ \hat{G}%
^{-1}_{EX}\left(P_i + \varepsilon, Y_i,Q_i\right) \leq \tau \right] K\left(%
\frac{W_i-w}{h}\right)K\left(\frac{Y_i-y}{h} \right) \\
= & \frac{1}{h^2} P_n \left\{ I\left[\hat{G}^{-1}_{EX}\left(P + \varepsilon,
Y,Q\right) \leq \tau \right]K\left(\frac{W-w}{h}\right)K\left(\frac{Y-y}{h}
\right) \right\}.
\end{align*}
Define  
\begin{align*}
R_{n1}(y,w,\varepsilon) = h^{-2} \left(P_n-P\right) %
        & \left\{ \left( I\left[ \hat{G}%
^{-1}_{EX}\left(P + \varepsilon, Y,Q\right) \leq \tau \right] - I\left[ {G}^{-1}_{EX}\left(P +
\varepsilon, Y,Q\right) \leq \tau \right]\right)\right. \\
        & \left. %
          K\left(\frac{W-w}{h}\right)%
          K\left(\frac{Y-y}{h} \right) \right\} \\
\end{align*}
and
\begin{align*}
R_{n2}(y,w,\varepsilon) = h^{-2} P \left\{ I\left[ \hat{G}^{-1}_{EX}\left(P
+ \varepsilon, Y,Q\right) \leq \tau \right] K\left(\frac{W-w}{h}\right)K\left(%
\frac{Y-y}{h} \right) \right\} \\
+ h^{-2}\left(P_n-P\right) \left\{I\left[ {G}^{-1}_{EX}\left(P +
\varepsilon, Y,Q\right) \leq \tau \right] K\left(\frac{W-w}{h}\right)K\left(%
\frac{Y-y}{h} \right) \right\}.
\end{align*}
Then $R_n = R_{n1} + R_{n2}$. In linear functional notation, $\hat{G}%
^{-1}_{EX}$ is treated as a fixed (non-random) function in the integrals.

Under Assumption 1, $\hat{G}^{-1}_{EX} - G^{-1}_{EX} =
O_p\left(n^{-1/2}\right)$. Therefore, it follows from Lemma 2.37 of \citet{Pollard1984} that  
\begin{align*}
R_{n1}(y,w,\varepsilon)=O_p\left[\frac{(\log n)^\gamma}{nh}\right]
\end{align*}
uniformly over $(y,w,\varepsilon)$. It further follows
that  
\begin{align*}
& R_n(y,w,\varepsilon)=h^{-2} P\left\{ I\left[\hat{G}_{EX}^{-1}
(P+\varepsilon,Y,Q) \leq \tau\right] K\left(\frac{W-w}{h}\right)K\left(\frac{Y-y%
}{h} \right) \right\} \\
& \;\;\;\;\;\; + h^{-2} (P_n-P) \left\{I\left[{G}_{EX}^{-1} (P+\varepsilon,Y,Q)
\leq \tau\right] K\left(\frac{W-w}{h}\right)K\left(\frac{Y-y}{h} \right)
\right\} + O_p\left[ \frac{(\log n)^\gamma}{nh}\right] \\
& = h^{-2} P\left\{ I\left[\hat{G}_{EX}^{-1} (P+\varepsilon,Y,Q) \leq \tau%
\right] - I\left[{G}_{EX}^{-1} (P+\varepsilon,Y,Q) \leq \tau\right]
\right\}K\left(\frac{W-w}{h}\right)K\left(\frac{Y-y}{h} \right) \\
& \;\;\;\;\;\; + h^{-2} P_n\left\{ I\left[{G}_{EX}^{-1} (P+\varepsilon,Y,Q) \leq
\tau\right] K\left(\frac{W-w}{h}\right)K\left(\frac{Y-y}{h} \right) \right\} +
O_p\left[ \frac{(\log n)^\gamma}{nh}\right] \\
& \equiv \; R_{n3} (y,w,\varepsilon) + R_{n4}(y,w,\varepsilon) + O_p\left[ 
\frac{(\log n)^\gamma}{nh}\right].
\end{align*}

Under Assumption 1, $(\hat\theta - \theta_0) = O_p(n^{-1/2})$. It follows from standard arguments for kernel estimators
that $R_{n3}(y,w,\varepsilon)=O_p(n^{-1/2})$ uniformly over $(y,w,\varepsilon)$. Therefore, by Assumption 2,
\begin{align}
        R_n(y,w,\varepsilon)=R_{n4}(y,w,\varepsilon) + O_p(n^{-1/2})\label{eqn4}
\end{align}
uniformly over $(y,w,\varepsilon)$.

Now consider $R_{n4}(y,w,\varepsilon)$. Because $U=G_{EX}^{-1}(P+%
\varepsilon,Y,Q)$,  
\begin{align*}
& R_{n4}(y,w,\varepsilon)=h^{-2} P_n \left[ I(U\leq \tau) K\left(\frac{W-w}{h}%
\right)K\left(\frac{Y-y}{h} \right) \right], \\
& R_{n4}(y,w,\varepsilon) - \tau \hat{f}_{YW}(y,w) = h^{-2} P_n \left\{\left[%
I(U\leq \tau) - \tau \right] K\left(\frac{W-w}{h}\right)K\left(\frac{Y-y}{h}
\right) \right\},
\end{align*}
and  
\begin{align}
        P\left[R_{n4}(y,w,\varepsilon) - \tau \hat{f}_{YW}(y,w) \right]=0.\label{eqn5}
\end{align}
Therefore, $R_{n4}(y,w,\varepsilon) -\tau \hat{f}_{YW}(y,w)$ %
is a mean-zero stochastic process. The covariance function of this process
is $[C(y_1, w_1; y_2, w_2) + o(1)]/(nh^2)$, where  
\begin{align*}
C(y_1,w_1; y_2,w_2) = \tau(1-\tau) f_{YW}(y_1,w_1) \int K(\xi) K(\xi +
\delta_W) K(\zeta) K(\zeta + \delta_Y) d\xi d\zeta, 
\label{feb27-eqn5}
\end{align*}
where $\delta_W=(w_1-w_2)/h$ and $\delta_Y = (y_1-y_2)/h$. It follows from (\ref%
{eqn4}) and (\ref{eqn5}) that  
\begin{align*}
        S_n(y,w) - \tau \hat{f}_{YW}(y,w) = \frac{1}{h^2} P_n \left\{\left[I(U\leq \tau) - \tau %
\right] K\left(\frac{W-w}{h}\right)K\left(\frac{Y-y}{h} \right) \right\} +
O_p\left(n^{-1/2}\right).
\end{align*}
Define the stochastic process  
\begin{align*}
Z_n(y,w) & = n^{1/2}h^{-1} P_n \left\{\left[I(U\leq \tau) - \tau \right]K\left(%
\frac{W-w}{h}\right)K\left(\frac{Y-y}{h} \right) \right\} \\
& = \frac{1}{n^{1/2}} h^{-1} \sum_{i=1}^n \left[I(U_i \leq \tau) - \tau \right]
K\left(\frac{W_i-w}{h}\right)K\left(\frac{Y_i-y}{h} \right) \\
& = n^{1/2} h [S_n(y,w) - \tau \hat f_{YW}(y,w)] + O_p(h).
\end{align*}
Let $\left\{\psi_j: j=1,2, \dots \right\}$ be the eigenfunctions of $C(y_1, w_1; y_2, w_2)$ and $\{\lambda_{nj}: j=1,2, ...\}$ be the eigenvalues. The $\psi_j$'s form a complete, orthonormal basis for $L_2[-1,1]^2$. $Z_n(y,w)$ has the representation 
\begin{align*}
        Z_n(y,w) & = \sum_{k=1}^\infty \hat b_{nk} \psi_k(y,w)
\end{align*}
where
\begin{align*}
        \hat{b}_{nk} & = \int Z_n(y,w)\psi_k(y,w) dy dw.
\end{align*}
Moreover,
\begin{align*}
        E\hat b_{nk} = 0
\end{align*}
and
\begin{align*}
        E(\hat b_{nk} \hat b_{nl})= \lambda_{nk}\delta_{kl} + o(1)
\end{align*}
for all $k$ and $l$, where $\delta_{kl}$ is the Kronecker delta. In addition,
\begin{align*}
        T_n & =\sum_{k=1}^\infty \hat{b}_{nk}^2.
\end{align*}
Let $\left\{L_n: n=1,2, \dots\right\}$ be an increasing sequence of positive
constants such that $L_n \rightarrow \infty$ as $n \rightarrow \infty$.
Define  
\begin{align*}
        \tilde T_n & =\sum_{k=1}^{L_{n}} \hat{b}_{nk}^2.
\end{align*}
Then  
\begin{align*}
|\tilde{T}_n - T_n | \rightarrow^p 0.
\end{align*}
Let $V_{L_n}$ denote the $L_n \times L_n$ diagonal matrix whose $(l,l)$ element is $\lambda_{nl}$. Let $\omega$ be a $L_n \times 1$ random vector with the $N(0,V_{L_n})$ distribution, and let $\norm*{\cdot}$ denote the Euclidean norm. It follows from Theorem A.1 of \citet{Spokoiny-Zhilova2015} that for any $z>max(4,L_n)$ and some constant $C_4<\infty$,
\begin{align*}
\abs*{\ P\left(\tilde{T}_n \leq z \right) - P\left(\norm*{\omega}^2 \leq z
\right)} \leq C_4 n^{-1/2} L_n^{3/2}.
\end{align*}
Assume that $n^{-1/2}L_n^{3/2} \rightarrow 0$ as $n \rightarrow \infty$.
Then  
\begin{align*}
P\left( T_n \leq z\right) - P\left( \norm*{\omega}^2 \leq z\right)
\rightarrow 0
\end{align*}
as $n \rightarrow \infty$, and the distribution of $T_n$ can be approximated
by that of $\norm*{\omega}^2$. This is  
\begin{align*}
        \norm*{\omega}^2 = \sum_{j=1}^{L_n} \lambda_{nj} \chi_j^2,
\end{align*}
where the $\chi_j^2\,$s are independent random variables that are distributed as chi-square with
one degree of freedom. Estimate the $\lambda_{nj}$'s by the eigenvalues of the empirical covariance operator of $Z_n$.

\end{document}